\documentclass[prd,showpacs,nofootinbib,preprintnumbers,
11pt ]{revtex4}
\usepackage{amsmath} \usepackage{graphicx} \usepackage{amsfonts}
\usepackage{array} \usepackage{amsthm} \usepackage{bm} \usepackage{mathptmx}

\usepackage{latexsym}
\newcommand{\nn}{\nonumber}
\newcommand{\be}{\begin{equation}}
\newcommand{\ee}{\end{equation}}
\newcommand{\ba}{\begin{eqnarray}}
\newcommand{\ea}{\end{eqnarray}}
\newcommand{\bal}{\begin{align}}
\newcommand{\eal}{\end{align}}
\newcommand{\lb}{\label}
\newcommand{\ovl}{\overline}
\newcommand{\e}{{\rm e}}
\newcommand{\dd}{{\rm d}}
\newcommand{\ii}{{\rm i}}

\newcommand{\om}{\omega}
\newcommand{\al}{\alpha}
\newcommand{\la}{\lambda}
\newcommand{\bt}{\beta}
\newcommand{\ka}{\kappa}
\newcommand{\pa}{\partial}

\newcommand{\si}{\sigma}

\newcommand{\fr}{\frac}

\newcommand{\vt}{\vartheta}
\newcommand{\tr}{{\rm tr\,}}
\newcommand{\bw}{\begin{widetext}}
\newcommand{\ew}{\end{widetext}}
\newcommand{\eps}{\epsilon}

\def\m{\pmb{\mu}}
\def\n{\pmb{\nu}}
\def\b{\pmb{b}}

\def\is{\frac1{\sqrt2}}
\newcommand{\cM}{{\cal M}}
\newcommand{\cN}{{\cal N}}
\newcommand{\cR}{{\cal R}}
\newcommand{\cE}{{\cal E}}
\newcommand{\cF}{{\cal F}}
\newcommand{\hA}{{\hat A}}
\newcommand{\hF}{{\hat F}}
\newcommand{\rr}{{\bf  r}}
\newcommand{\R}{\mathbb{R}}
\renewcommand{\theequation}{\arabic{section}.\arabic{equation}}

\begin{document}
\begin{flushright}LAPTH-027/11
\end{flushright}
\begin{flushright}DTP-MSU/11-12
\end{flushright}

\title{All extremal instantons in Einstein-Maxwell-dilaton-axion theory}
\author{Mustapha Azreg-A\"{\i}nou}\email{azreg@baskent.edu.tr}
\affiliation {Ba\c{s}kent University, Department of Mathematics,
Ba\u{g}l\i ca Campus, Ankara, Turkey}
\author
{G\'erard Cl\'ement}\email{gclement@lapp.in2p3.fr} \affiliation{
Laboratoire de  Physique Th\'eorique LAPTH (CNRS),
\\
B.P.110, F-74941 Annecy-le-Vieux cedex, France}
\author{Dmitri V. Gal'tsov} \email{galtsov@phys.msu.ru}
\affiliation{Department of Theoretical Physics, Moscow State
University, 119899, Moscow, Russia}

\date{\today}

\begin{abstract}
We construct explicitly all extremal instanton solutions to
$\cN=4,\, D=4$ supergravity truncated to one vector field
(Einstein-Maxwell-dilaton-axion (EMDA) theory). These correspond to
null geodesics of the target space of the sigma-model
$G/H=Sp(4,\mathbb{R})/GL(2,\R)$ obtained by compactification of
four-dimensional Euclidean  EMDA on a circle. They satisfy a
no-force condition in terms of the asymptotic charges and part of
them (corresponding to nilpotent orbits of the $Sp(4,\mathbb{R})$
U-duality) are presumably supersymmetric. The space of finite action
solutions is found to be unexpectedly large and includes, besides
the Euclidean versions of known Lorentzian solutions, a number of
new asymptotically locally flat (ALF) instantons endowed with
electric, magnetic, dilaton and axion charges. We also describe new
classes of charged asymptotically locally Euclidean (ALE) instantons
as well as some exceptional solutions. Our classification scheme is
based on the algebraic classification of matrix generators according
to their rank, according to the nature of the charge vectors  and
according to the number of independent harmonic functions with
unequal charges. Besides the nilpotent orbits of $G$, we find
solutions which satisfy the asymptotic no-force condition, but are
not supersymmetric. The renormalized on-shell action for instantons
is calculated using the method of matched background subtraction.
\end{abstract}

%% REVTEX4

\maketitle

\section{Introduction}

Gravitational instantons are important ingredients of quantum
gravity/supergravity and string theory  responsible for their
non-perturbative aspects. They may produce a non-trivial topological
structure of space-time in the early universe, and play certain role
in cosmology. Gravitational instantons exhibiting periodicity in
Euclidean time are stationary paths for thermal partition functions,
these are responsible for black hole thermodynamics.

Instantons in vacuum Einstein gravity were subject of intense
investigations since the late seventies \cite{inst, Gibbons:1979nf},
which culminated in their complete topological classification
\cite{Eguchi:1980jx}. Instantons in extended supergravities are
non-vacuum and typically involve multiplets of scalar and vector
fields in four dimensions and form fields in higher dimensions.
Non-vacuum gravitational instantons attracted attention in the late
eighties, when they were suggested to support the idea of fixing the
physical constants via creation of baby universes \cite{baby}. Such
instantons and Euclidean wormholes were studied within truncated
supergravity models containing the axion and the dilaton. In this
context, Einstein-axion \cite{axion} and Einstein-dilaton-axion
\cite{dilaton-axion} wormhole solutions were discovered. These are
particular solutions of the Einstein-Maxwell-dilaton-axion theory
\cite{Kallosh:1992wa} which is here the subject of more complete
investigation. Essentially similar solutions, D-instantons, were
then discovered in ten-dimensional IIB supergravity and generalized
to any dimensions \cite{Dinstanton}. Wormhole instanton solutions
also exist in four-dimensional Einstein-Maxwell-dilaton theory
(without axion) \cite{caca}. Non-vacuum gravitational instantons
were also explored in the presence of a cosmological constant, in
which case they exhibit a de Sitter or anti-de Sitter asymptotic
structure \cite{Gutperle:2002km}. Asymptotically AdS wormholes
provide an arena for further study of the ADS/CFT correspondence,
which in turn may be used to test the validity of the above
proposals of wormhole-induced effects \cite{arkani}. Apart from
various applications directly related to quantum gravity,
four-dimensional instantons can be useful in the purely classical
theory as a tool to construct new five-dimensional black holes
\cite{Bena:2009fi}.

During the past two decades, instantons including vector fields were
extensively studied in $D=4, \cN=2$ supergravity. An early work by
Whitt \cite{whitt} and Yuille \cite{yuil} discussing instantons in
Einstein-Maxwell gravity was recently revised and extended by
Dunajski and Hartnoll \cite{Dunajski:2006vs} (for a
higher-dimensional extension see \cite{Awad:2005ff}). These papers
deal with the Euclidean counterparts of the Israel-Wilson-Perj\`es
solutions \cite{IWP}. Euclidean solutions to $\cN=2$ supergravity
coupled to various matter multiplets were recently studied in detail
in a number of papers \cite{Chiodaroli:2009cz}.   General aspects of
supersymmetry and dualities in Euclidean supergravities were
discussed in \cite{Cremmer:1998em}. Euclidean supersymmetry and
Killing spinor equations in the Euclidean $\cN=2$ theory were
recently studied with \cite{Dunajski:2010zp} and without
 a cosmological constant \cite{Gutowski:2010zs}. Instantons in $D=4, \cN=1$ supergravity were
studied in \cite{Huebscher:2009bp}. In the $\cN=4$ case the complete
Killing spinor analysis   is available in the Lorentzian sector
\cite{Tod:1995jf,Bellorin:2005zc}.

The case of Euclidean $D=4,\, \cN=4$ supergravity in the presence of
vector fields was relatively less  explored so far, though uncharged
axion-dilatonic instantons and wormholes were discovered long ago.
In the minimal case this theory contains six vector fields
stransforming under $SO(6)$.  Here we consider truncation of $\cN=4$
theory to the EMDA theory with only one vector present, leaving the
full theory to further work. With this simplification we will be
able to give explicitly all extremal instanton solutions, whose
variety turns out to be unexpectedly large already in this truncated
case. We define four-dimensional extremal instantons as those which
have flat three-dimensional slices. Such solutions are characterized
by asymptotic charges: the mass $M$, the NUT parameter $N$, the
electric $Q$ and magnetic $P$ charges, the dilaton charge $D$ and
the axion charge $A$. In the Euclidean theory, the mass, the dilaton
charge and the electric charge generate attraction, while the
magnetic mass (NUT), the axion and the magnetic charge generate
repulsion. Extremality corresponds to fulfilment of the ``no
force'' condition
 \be\label{nof}
M^2+D^2+Q^2=N^2+A^2+P^2,
 \ee which is part of the BPS conditions of
$D=4,\, \cN=4$ supergravity, but does not guarantee supersymmetry,
being only the necessary condition for it. We do not investigate the
Killing spinor equations here, also leaving this to future work, but
we believe that all  supersymmetric solutions to the one-vector
truncation of $D=4,\, \cN=4$ supergravity belong to our list.

Here we use purely bosonic tools to identify solutions satisfying
the no-force condition (\ref{nof}). The method amounts to identify
the null geodesic curves of the target space of the sigma model
arising upon dimensional reduction of the theory to three
dimensions. The underlying heuristic idea relates to the fact that
by virtue of the Einstein equations, null-geodesic solutions have
Ricci-flat three-metrics which are therefore flat since the Weyl
tensor is zero in three dimensions. The method was suggested by one
of the present authors in 1986 \cite{gc}, building on the
characterization by Neugebauer and Kramer \cite{nk} of solutions
depending on a single potential as geodesics of the
three-dimensional target space, in the context of five-dimensional
Kaluza-Klein (KK) theory. It was further applied in \cite{bps} to
classify Lorentzian extremal solutions of the EMDA theory and
Einstein-Maxwell (EM) theory. In two of these three cases (KK and
EMDA) it was found that the matrix generators $B$ of null geodesics
split into  a nilpotent class ($B^n=0$ starting with certain integer
$n$), in which cases the matrix is degenerate ($\det B=0$), and a
non-degenerate class ($\det B\neq 0$). The solutions belonging to
the first class are regular, while the second class solutions,
though still satisfying the no-force constraint (\ref{nof})  on
asymptotic charges, generically contain singularities. In the EM
case all null geodesics are nilpotent orbits, corresponding to the
Israel-Wilson-Perj\`es solutions \cite{bps}.

This approach partially overlaps (though a bit wider) with the
method of  nilpotent orbits which was suggested in
\cite{Gunaydin:2005mx} (see also \cite{Gunaydin:2009pk}) and further
developed in \cite{Boss}. The latter starts with some matrix
condition following from supersymmetry, which is generically
stronger than our condition selecting the null geodesic subspace of
the target space. Our classification includes some a priori
non-supersymmetric solutions satisfying the no-force condition
(\ref{nof}).

The purpose of this paper is to give a complete list of null
geodesic instantons of EMDA theory saturating the asymptotic bound
(\ref{nof}).  Technically, we will use the Euclidean version of the
EMDA sigma model derived in \cite{emda} and equipped with a concise
symplectic matrix representation in \cite{diak}. In the Lorentzian
case, the six-dimensional symmetric target space of the sigma-model
obtained by time-like dimensional reduction (appropriate to the
generation of black hole solutions) of four-dimensional EMDA is
$Sp(4,R)/U(1,1)$, while in the case of space-like reduction it is
$Sp(4,R)/U(2)$. In the Euclidean case one finds yet another coset of
the same dimensionality six, namely $Sp(4,R)/GL(2,R)$.  Our results
indicate that the EMDA instanton space is much larger than one could
anticipate using analytical continuation of the known Lorentzian
solutions. The target space geometry of Euclidean EMDA differs by
signature from that of the Lorentzian theory, and contains three
independent null directions compared to only two in the Lorentzian
case. This gives rise to new classes of three-potential extremal
solutions. Also, new classes of solutions arise which are not
asymptotically locally flat (ALF) but asymptotically locally
Euclidean (ALE). Finally, there are exceptional solutions for which
the dilaton diverges at infinity, while the renormalized action is
still finite. Generically, the on-shell action for four-dimensional
gravitating instantons always diverges at infinity due to slow
fall-off of the Gibbons-Hawking term, so it has to be renormalized.
As the renormalization tool, we adopt the matched asymptotic
subtraction method.

The outline of the paper is as follows. The Euclidean
four-dimensional Einstein-three-form-Maxwell-dilaton action is
discussed in Sect. II. We perform in Sect. III toroidal reduction of
this theory to three dimensions, keeping track of the boundary terms
arising in the dualizations involved, which will be relevant for the
calculation of the instanton on-shell actions. This reduction leads
to a three-dimensional gravitating sigma model with symmetric target
space $Sp(4,\mathbb{R})/GL(2,\R)$ of signature ($+++---$). In Sect.
IV we construct the matrix representation of this coset and discuss
the different asymptotic forms of the coset matrices, leading to ALF
and ALE instanton solutions. Some exceptional asymptotics are
discovered within each class. In Sect. V we introduce null geodesic
solutions. The associated matrix representatives are parametrized in
terms of asymptotic charges saturating the Bogomolny bound. We
derive a simple formula for the instanton action involving only the
boundary values of the scale factor and the dilaton function and
their derivatives. We also give a convenient description of the
matrix generators in terms of $SO(1,2)$ charge vectors, which will
provide the basis for further classification of solutions.  Sect. VI
presents the classification of ALF instantons which are split into
strongly degenerate (nilpotent of rank 2), weakly degenerate
(nilpotent of rank 3) and  non-degenerate. The corresponding three
kinds of ALE solutions are discussed in Sec. VII. One of these includes
a new types of wormhole interpolating between ALE and conical
ALF spaces. Examples of ALF and ALE instantons with exceptional
asymptotics, including a magnetic linear dilaton solution, are also given.
In Sec. VIII we
consider the case of multiple independent harmonic functions  whose
maximal number (three) is determined by the number of independent
null directions in target space. The last Sec. IX is devoted
to a brief discussion of the six-dimensional uplifting of
four-dimensional EMDA instantons.  In Appendix A we briefly discuss the
sigma-model representation of the ``phantom'' (with a Maxwell field coupled
repulsively to gravity) EMDA model with positive definite signature of the
target space. Details of the derivation of solutions with exceptional
asymptotics are given in Appendix B, and the proof that multi-potential
solutions fall into three distinct classes is given in Appendix C.

\setcounter{equation}{0}
\section{Euclidean EMDA theory}
Recall that the correct choice of the Euclidean action for the axion
field in four dimensions follows from  positivity requirement and
amounts to using initially the three-form field rather than a
pseudoscalar axion. The corresponding action with account of  the
Gibbons-Hawking surface term reads
\begin{equation}\label{acH}
    S_0=\frac1{16\pi}\int\limits_{\cM}   \left(
    -R\star 1+2  \dd\phi\wedge \star
\dd\phi+  2\e^{-4\phi}  H\wedge  \star H + 2\e^{-2\phi}
 F\wedge \star  F \right)\; - \frac1{8\pi}\int\limits_{\pa\cM}\e^{\psi/2}K\star\dd \Phi\,,
\end{equation}
 where $F=\dd A$ is the Maxwell two-form and $H$ is the
three-form field strength related to the two-form potential $B$ via
the relation involving the Chern-Simons term:
 \be
H=\dd B-A\wedge F.
 \ee
The four-dimensional Hodge dual is denoted by a star $\star$ and
defined in local coordinates by
\begin{equation}\label{2h0}
    \star(\dd x^{\mu_1}\wedge\cdots\wedge\dd x^{\mu_m})=
    \frac{1}{(4-m)!}\,E^{\mu_1\cdots\mu_m}{}_{\mu_{m+1}\cdots\mu_4}
    \dd x^{\mu_{m+1}}\wedge\cdots\wedge\dd x^{\mu_4}\,,\;m\leq 4\,,
\end{equation}
where the totally anti-symmetric symbol $\epsilon_{\mu\nu\rho\si}$
and tensor $E_{\mu\nu\rho\si}$ are related by
$E_{\mu\nu\rho\si}=g^{1/2}\epsilon_{\mu\nu\rho\si}$,
$E^{\mu\nu\rho\si}=g^{-1/2}\epsilon^{\mu\nu\rho\si}$, with
$\epsilon_{1234}=\epsilon^{1234}=+1$. The boundary $\pa\cM$, which
is embedded in $\cM$, is described by $\Phi(x^{\mu})\equiv 0$, while
$\e^{\psi/2}$ is a scale factor ensuring that $\e^{\psi/2}\dd \Phi$
measures the proper distance in a direction normal to $\pa\cM$, and
$K$ is the trace of the extrinsic curvature of $\pa\cM$ in $\cM$.

The corresponding field equations are
 \begin{align}
 &\dd\star \left[\e^{-4\phi}\left(\dd B-A\wedge F\right)\right]=0,\\
 & \dd\star\left(\e^{-2\phi}F\right)=0,\\
 & \square \phi=\fr14\,\e^{-2\phi}F^2+\fr16\,\e^{-4\phi}H^2.
 \end{align}
It follows that $A=0$ is a consistent truncation of EMDA which is
pure dilaton-axion gravity, while it is not consistent to set $B=0$
and $\phi=0$ (constraints on $F$ will be produced), so
Einstein-Maxwell theory is not a consistent truncation of EMDA. Of
course this does not preclude the possibility of EMDA solutions with
zero dilaton and axion fields, which satisfy the arising constraints
on the vector field.

To pass to the pseudoscalar axion consistently, one has to ensure
the Bianchi identity for $H$:
 \be\lb{bianH}
  \dd \dd B=\dd (H + A\wedge F)=0\,.
 \ee
This is achieved adding to the action (\ref{acH}) a new term with
the Lagrange multiplier $\ka$
 \be
S_{\kappa}= \frac1{8\pi}\int_{\cM'} \ka\; \dd (H + A\wedge F)=
\frac1{8\pi}\int_{\cM'}\ka \; (\dd H + F\wedge F)\,,
 \ee
where $\cM'$ is $\cM$ with the monopole sources of $H$ (where the
Bianchi identity (\ref{bianH}) breaks down) cut out. In what
follows, we will use both the open set $\cM'$ and the original
manifold $\cM$ as the integration domains, keeping in mind that in
absence of magnetic singularities integrals of both types will be
the same. Transforming the first term as
 \be   \int_{\cM'} \ka \dd  H = \int_{\cM'}
\dd\; \left(\ka  H  \right)-\int_{\cM'}  \dd\ka \wedge H,
 \ee
we obtain a boundary term from the total derivative, while the
$H$-dependent part of the bulk action will be
 \be
S_H= \frac1{8\pi}\int   \left(\e^{-4\phi}H\wedge \star H- \dd\ka
\wedge H\right).
 \ee
Treating $H$ as fundamental field rather than the $B$-field
strength, we obtain varying $S_H$ over $H$: \be H=\frac12\,
\e^{4\phi} \star \dd \ka. \ee Eliminating $H$ in favor of $\ka$ then
gives
 \be\lb{SHk}
S_H= -\frac1{32\pi}\int   \e^{4\phi} \dd\ka \wedge \star \dd\ka.
 \ee
Thus the sum $S_0+S_{\kappa}$ gives the Euclidean bulk action  in terms of
the pseudoscalar axion
\begin{equation}\label{ac2}
    S_E= \frac1{16\pi}\int\limits_{\cM}   \left(
    -R\star 1+2  \dd\phi\wedge
\star\dd\phi- \fr12\, \e^{4\phi}  \dd \ka\wedge \star\dd\ka  +
2\e^{-2\phi}
 F\wedge  \star F + 2\ka F\wedge F\right)\,,
\end{equation}
or in the component form
\begin{equation}\label{ac3}
S_E= \frac1{16\pi}\int \dd^4x\sqrt{g}\left[
-R+2\partial_{\mu}\phi\partial^{\mu}\phi-
\frac{1}{2}\,\e^{4\phi}\partial_{\mu}\kappa\partial^{\mu}\kappa
+\e^{-2\phi}F_{\mu\nu}F^{\mu\nu} -\kappa
F_{\mu\nu}\tilde{F}^{\mu\nu}\right]\,.
\end{equation}
where $\tilde{F}^{\mu\nu}=-E^{\mu\nu\rho\si}F_{\rho\si}/2$ (the
unusual minus sign is due to our convention $\epsilon_{1234}=+1$,
which is usually taken as $\epsilon_{0123}=+1$).
The last term in (\ref{ac3}) coupling the axion to the pseudoscalar
Maxwell invariant does not depend on the metric, so it does not
contribute to the Einstein equations:
\begin{equation}\lb{einst4}
R_{\mu\nu}=2\pa_\mu\phi\pa_\nu\phi -
\frac{1}{2}\,\e^{4\phi}\pa_\mu\kappa \pa_\nu\kappa+
\e^{-2\phi}(2F_{\lambda\mu}{F^{\lambda}}_\nu -
\frac{1}{2}\,F_{\lambda\tau}F^{\lambda\tau}g_{\mu\nu})\,.
\end{equation}

Our purpose will be to solve the equations of motion for extremal
instantons and to calculate the on-shell action for them. However,
the latter is usually divergent. The boundary manifold ${\pa\cM}$
will generically consist of an external boundary ${\pa\cM}_r$
corresponding to some finite value of a suitably chosen radial
coordinate $r$, and an inner boundary ${\pa\cM}_{\rm int}$ which must be
introduced if solution under consideration has curvature
singularities. In this paper we will consider ALF and/or ALE
solutions and send the external boundary ${\pa\cM}_r$ to infinity
($r \to\infty$) at the end of the calculation. It is well-known that
already in absence of the inner boundary, such a calculation is
ambiguous since in four space-time dimensions the gravitational
contribution generically diverges in the limit $r \to\infty$. The
matter part of the action can also diverge (see Sects. IV and VII).
This problem was encountered in many applications related to
gravitational instantons and black hole thermodynamics which
corresponds to black hole solutions periodic in imaginary time.

Two main tools were suggested to overcome this difficulty. One
simple way to renormalize the action consists in subtracting the
(also infinite) action calculated for a reference solution
(background) which must be matched with the solution in question on
the boundary. This leads to a finite action, but the result will
depend on the particular chosen background. Moreover, it is not
always possible to embed the boundary geometry in the background
space-time exactly, so one has to resort to an approximate
embedding. The matching procedure then becomes non-trivial, as
revealed the discussion of the late nineties
\cite{Hunter:1998qe,Hawking:1998jf,Hawking:1998ct,Chamblin:1998pz}
of the action for the Taub-NUT and Taub-bolt instantons given
earlier by Gibbons and Perry \cite{Gibbons:1979nf}. While direct
omission of the divergent term in the action led to the values
$S_{TN}=4\pi N^2$ and $S_{Tb}=5\pi N^2$ for the self-dual Taub-NUT
instanton and the Taub-bolt \cite{Gibbons:1979nf}, the choice of
self-dual Taub-NUT as background leads respectively to $S_{TN}=0$
and $S_{Tb}=\pi N^2$ \cite{Hunter:1998qe}.

Both of the above values for the Taub-NUT instanton action can be
given physical meaning \cite{Kraus:1999di,Solodukhin:1999zr} within
the second regularization scheme for gravity-coupled theories which
was first suggested \cite{Balasubramanian:1999re,Emparan:1999pm}, in
the context of AdS/CFT correspondence, and later developed into the
holographic renormalization technique \cite{de
Haro:2000xn,Skenderis:2002wp}. This method consists in adding to the
$D$-dimensional bulk action a ($D-1$)-dimensional counterterm
action, depending uniquely on the boundary geometry, which cancels
the divergences of the on-shell gravitational action, background
subtraction being then unnecessary
\cite{Balasubramanian:1999re,Kraus:1999di}. This proposal was first
formulated in the case of AdS asymptotics
\cite{Balasubramanian:1999re}, where the corresponding boundary
stress-tensor has an  interpretation as the vacuum expectation value
of the stress-tensor operator of the quantum field theory
holographically dual to the bulk gravity-coupled theory. It has
since been realized that a similar procedure remains valid in the
asymptotically flat limit of an infinite curvature radius of the
AdS, and different proposals for counterterm actions were made
\cite{Kraus:1999di,Solodukhin:1999zr,Lau:1999dp,Mann:1999pc}, which
all cancel divergences. Moreover, the asymptotically flat case may
also admit a holographic interpretation
\cite{Polchinski:1999ry,Susskind:1998vk} in the context of M-theory.
The advantage of the method of geometric counterterms (apart from
its direct relation to holography) is that it removes divergences
without subtraction, and is thus independent of the choice of a
reference solution. This method ascribes to the vacuum self-dual
Taub-NUT instanton the non-zero Gibbons-Perry value of the action
$S_{TN}=4\pi N^2$ \cite{Emparan:1999pm,Solodukhin:1999zr,
Lau:1999dp,Mann:1999pc}. Accordingly, the holographically
renormalized action for the Taub-bolt instanton also corresponds to
the ``naive'' subtraction of the Minkowskian value of the divergent
term. However, the drawback of the holographic renormalization of the
action in the case of ALF and/or ALE Euclidean instantons is that
the general procedure suggested for generating counterterms is not
unique, and the finite part remaining after cancellation of
divergences is therefore also ambiguous \cite{Kraus:1999di,
Solodukhin:1999zr,Lau:1999dp,Mann:1999pc,Cai:1999xg}. Another
drawback is that the result may depend on the choice of coordinates
\cite{Kraus:1999di,Solodukhin:1999zr}.

In this paper we evaluate in a first step the actions for different
extremal instantons, without discussing possible applications to
quantum gravity/holography. For this purpose, we shall adopt the
conceptually and technically simpler procedure of Hunter
\cite{Hunter:1998qe} which prescribes subtraction of a suitably
matched vacuum background solution, which will be self-dual Taub-NUT
in the case of ALF solutions. It will therefore give by definition
the value zero for the vacuum Taub-NUT instantons. Other instantons
endowed with electric, magnetic and scalar charges will lead to a
non-zero action, these may be regarded as excitations of vacuum
Taub-NUT, similarly to the Taub-bolt instanton in the pure gravity
case.

Since the bulk term in (\ref{ac2}) is zero for the reference
space-time, to renormalize the action it will be enough to subtract
the matched background integrals over ${\pa\cM}_r$ in the
Gibbons-Hawking term and the potentially divergent axion boundary term
replacing $K$ by $[K]=K-K_0$, and $\ka \e^{4\phi}\star \dd\ka $ by
$[\ka \e^{4\phi}\star \dd\ka]= \ka \e^{4\phi}\star \dd\ka-\ka_0
\e^{4\phi_0}\star \dd\ka_0$, where $K_0$ is the reference value of
the trace of extrinsic curvature evaluated for an appropriate
background metric $g_{0}$ solving the field equations and matched to
the considered metric on the boundary $\pa\cM_r$, and
$\ka_0,\,\phi_0$ are analogous matter background terms. Collecting
then all the boundary terms, we obtain
 \be\lb{GHax}
\textsuperscript{4}S_b= \frac1{16\pi}\int\limits_{\pa\cM'}[\ka
\e^{4\phi}\star \dd\ka] -
\frac1{8\pi}\int\limits_{\pa\cM}\e^{\psi/2}[K]\star\dd \Phi,
 \ee
where the pull-back of the three-form $\star \dd\ka$ onto the
boundary $\pa\cM$ is understood. Note that the bulk matter action in
the form (\ref{ac3}) is not positive definite in contrast to
(\ref{acH}): the difference is hidden in the boundary term.

 Some of our new instantons contain inner singularities, for which
the above infrared renormalization is not sufficient and the
integrals still diverge in the UV. In such cases the instanton
action  will remain undetermined.

 \setcounter{equation}{0}
\section{Kaluza-Klein reduction with boundary terms}

To develop generating technique for instantons we apply dimensional
reduction to three dimensions. The derivation of the EMDA sigma
model was first given in \cite{emda} and further developed in
\cite{diak}. Its Euclidean version can be obtained by analytic
continuation, but we want in addition to keep boundary terms which
arise upon dualization of the Kaluza-Klein one-form and the
three-dimensional Maxwell one-form and which were disregarded in
earlier work. We therefore present here a more complete derivation.

Consider an oriented manifold $\cM$ with a Riemannian metric
$g_{\mu\nu}$ admitting an $U(1)$ isometry generated by the Killing
vector field $\xi=\pa_t$ where $t$ now is an Euclidean coordinate
with period $\beta$. The metric reads
\begin{equation}\label{an}
\dd s^2=g_{\mu\nu}\dd x^\mu \dd x^\nu=f(\dd t-\omega_i\dd
x^i)^2+\frac{1}{f}\,h_{ij}\dd x^i\dd x^j\,,
\end{equation}
where $f,\,\omega_i,\,h_{ij}$ are functions of $x^i$ ($i=1,2,3$).
Occasionally we will also use an exponential parametrization  of the
scale factor $f=\e^{-\psi}$. The classification of such metrics in
terms of the fixed point sets of $\xi$ was given  by Gibbons and
Hawking \cite{inst}. In four-dimensional space, the submanifolds on
which the norm $\xi^\mu\xi_\mu=f(x^i)$ vanishes may be
two-dimensional (bolts), or zero-dimensional (nuts). A regular
foliation of space in terms of the orbits of $\xi$ is possible on
the manifold $\cM'$ obtained from $\cM$ after subtracting the sets
of its fixed points. The contribution to the action of the nuts and
bolts can be computed in a general way using either Kaluza-Klein
reduction \cite{Hunter:1998qe}, or ADM decomposition with an
associated Hamiltonian formalism \cite{Hawking:1998jf}. Our
Lagrangian is not of the form considered in \cite{Hawking:1998jf},
so the results of this paper can not be directly applied here.
Rather we would need for this purpose the Euclidean version of the
Hamiltonian formulation of EMDA presented in \cite{Clement:2002mb}.
We postpone Hamiltonian analysis for a future paper, and here
restrict ourselves to Kaluza-Klein reduction to three dimensions,
with careful account for the boundary terms.

The dimensional reduction of the gravitational action is standard.
Since the $t$ coordinate parametrizes the circle of circumference
$\beta$, the integral over $\cM$ reduces to the integral over a
three-dimensional Euclidean space $\cE$ with two-dimensional
boundary $\pa\cE$ described by $\Psi(x^i)=\Phi(x^{\mu})|_{\pa\cE}\equiv
0$. Throughout this paper the three-dimensional Hodge
dual is denoted by an asterisk $\ast$ and defined by a similar
formula to~\eqref{2h0} where we replace $4$ by $3$ and $\sqrt{g}$ by
$\sqrt{h}$ and set the conventions
$\epsilon_{123}=\epsilon^{123}=+1$. Reducing $R$ \`a la Kaluza-Klein
we obtain
 \be \lb{gr}
-\frac1{16\pi}\int\limits_{\cM} R\star 1 = \frac{\beta}{16\pi}
\int\limits_{\cE}\left(-\cR\ast 1+
\fr12\ast\dd\psi\wedge\dd\psi +\fr12\e^{-2\psi}\ast\cF
\wedge\cF\right) + {}^3S_{b1}\,,
 \ee
where $\om=\om_i \dd x^i$ and $\cF=\dd \om$ are the Kaluza-Klein
one- and two-form, respectively. The boundary term due to the total
derivative reads
 \be\lb{b1}
{}^3S_{b1}=\frac{\beta}{16\pi}\int\limits_{\pa\cE}\ast\dd\psi\,.
 \ee
On the other hand, dimensional reduction of the regularized four-dimensional
Gibbons-Hawking term leads to
 \be\lb{redGH}
\textsuperscript{4}S_{GH}=-\frac{1}{8\pi}
\int\limits_{\pa\cM}\e^{\psi/2}[K]\star\dd \Phi =
-\frac{\beta}{16\pi}\int\limits_{\pa\cE} (2[k]\ast\dd \Psi +
\ast[\dd\psi])\,,
 \ee
where $[k]$ is the regularized trace of the extrinsic
curvature of $\pa\cE$ embedded in $\cE$. The last term partly
cancels the contribution from the reduction of the bulk gravity
(\ref{b1}), however there is a residual gravitational contribution to the
two-dimensional boundary term
 \be\lb{bG}
\textsuperscript{3}S_{bG} = \frac{\beta}{16\pi}\int\limits_{\pa\cE}
(-2[k]\ast\dd \Psi + \ast\dd\psi_0)\,.
 \ee

In the matter part of the action (\ref{ac2}) we have to keep in mind
the transgression rules. Equipping temporarily the four-dimensional
Maxwell potential and the  two-form strength with a hat,
$\dd\hF=\hA$, and denoting the corresponding three-dimensional forms
as $A,\,F$, we define (to simplify notation we will temporary use
electric and magnetic potentials differing from those in the
previous section by $\sqrt{2}$, they will be rescaled  back at the
end of the calculation)
 \be\lb{defv}
\hA=v\dd t +A,\qquad F=\dd v \wedge \om+\dd A\,,
 \ee
so that
 \be
\hF=\dd v \wedge \vartheta +F,\qquad \vartheta=\dd t-\om.
 \ee
Then the four-dimensional and the three-dimensional duals will be
related via
 \be
\star\hF=\e^\psi\ast \dd v  +\e^{-\psi}\ast F\wedge\vt\,.
 \ee
Substituting this into the Eq.~\eqref{ac2} and combining with
(\ref{gr}) we obtain the bulk action
 \begin{align}\lb{ac4}
S_{E}&=\frac{\beta}{16\pi}\int\limits_{\cE}\Big[-\cR\ast 1+
\fr12\ast\dd\psi\wedge\dd\psi +\fr12\,\e^{-2\psi}\ast\cF \wedge\cF+2
\ast \dd\phi\wedge \dd\phi \nonumber \\
&-\fr12\, \e^{4\phi} \ast \dd \ka\wedge
\dd\ka+2\e^{-2\phi}\left(\ast\dd v\wedge \dd v \e^\psi+ \ast F\wedge
F\e^{-\psi}\right)+4\ka \dd v\wedge F\Big]\,.
 \end{align}

Now we wish to dualize the three-dimensional Maxwell two-form
 \be\lb{3duF}
F=\ast g,
 \ee
 where $g$ is the one form $g=g_i \dd x^i$. To guarantee the
 validity of the Bianchi identity
 $\dd (F-\dd v \wedge \om) =0$
we add to the action (\ref{ac4}) a new term with Lagrange multiplier
$u$ (the magnetic potential)
 \be\lb{u}
S_u=\frac{\beta}{4\pi}\int\limits_{\cE'} u \;\dd (F-\dd v \wedge \om)=
\frac{\beta}{4\pi}\int\limits_{\cE'} \left[-\dd u \wedge F + u\dd v
\wedge\cF\right] +\frac{\beta}{4\pi}\int\limits_{\pa\cE'} uF\,.
 \ee
Now consider the last two terms in the action (\ref{ac4}) together
with the $\dd u \wedge F$ term in the volume part of (\ref{u}), and
perform variation over $g$ as a fundamental field. We obtain
 \be\lb{defu}
g = (\dd u -\ka\dd v)\,\e^{2\phi+\psi}\,.
 \ee
This can be now substituted back into the action (\ref{ac4}) to
eliminate the Maxwell three-dimensional two-form in favor of the
scalar magnetic potential $u$
 \begin{align}\lb{ac5}
S_{E}&=\frac{\beta}{16\pi}\int\limits_{\cE}\Big[-\cR\ast 1+
\fr12\ast\dd\psi\wedge\dd\psi +\fr12\,\e^{-2\psi}\ast\cF \wedge\cF+2
\ast \dd\phi\wedge \dd\phi - \fr12\, \e^{4\phi} \ast \dd
\ka\wedge \dd\ka \nonumber \\
&+2\e^{\psi-2\phi}\ast\dd v\wedge \dd v - 2\e^{\psi+2\phi}\ast (\dd
u -\ka\dd v)\wedge (\dd u -\ka\dd v) + 4u\dd v\wedge \cF\Big]\,.
 \end{align}

The last dualization is that of the Kaluza-Klein two-form. For this
we write $ \cF=\ast\varpi $ and add a new term, with Lagrange
multiplier $\eta$, ensuring the Bianchi identity $\dd \cF=0$:
 \be
S_{\eta}= \frac{\beta}{16\pi}\int\limits_{\cE'} \eta\dd\cF =
\frac{\beta}{16\pi}\int\limits_{\cE'} (-\dd\eta\wedge\cF) +
\frac{\beta}{16\pi}\int\limits_{\pa\cE'} \eta\cF\,.
 \ee
Collecting all $\cF$-terms in (\ref{ac5}) and performing variation
over $\varpi$ as a fundamental field, we obtain
 \be\lb{defchi}
\varpi=\e^{2\psi}(\dd\eta - 4u\dd v) = \e^{2\psi}(\dd\chi - 2u\dd v
+ 2v\dd u)\,, \quad (\chi \equiv \eta - 2uv )\,.
 \ee
Substituting this back in (\ref{ac5}) we eliminate $\cF$ from the
action replacing it with the NUT potential $\chi$.

To make contact with the notation of \cite{emda} we rescale the
electric and magnetic potentials $u\to u/\sqrt{2},\,v\to
v/\sqrt{2}\,$. The defining equations (\ref{defv}), (\ref{defu}) and
(\ref{defchi}) for $v,u,\chi$ become, in component notation
\begin{align} &F_{i4}=\frac{1}{\sqrt{2}}\,\partial_iv\,,\\&
\lb{mag} \e^{-2\phi}F^{ij}-\kappa {\tilde
F}^{ij}=\frac{f}{\sqrt{2h}}\,\epsilon^{ijk}
\partial_ku\,,\\
&\lb{twist}
\partial_i\chi +v\partial_iu-u\partial_iv=-f^2h_{ij}\,
\frac{\epsilon^{jkl}}{\sqrt{h}}\,\partial_k\omega_l
\end{align}
(here again the unusual minus sign in front of $\kappa$ is due to
our convention $\epsilon_{1234}=+1$). The full bulk action is that
of the gravity-coupled three-dimensional sigma model
 \be\lb{acsig}
S_\sigma = -\frac{\beta}{16\pi}\int \dd ^3x \sqrt{h}\left({\cal R}-
G_{AB}\partial_iX^A\partial_j X^B h^{ij}\right)\,\,,
\end{equation}
where the target space variables are $\textbf{X} =
(f,\phi,v,\chi,\kappa,u)\,,$ integration is over the three-space
$\cE$ and the target space metric $\dd l^2 = G_{AB}\dd X^A \dd X^B$
reads
 \be\lb{tar4} \dd l^2 = \frac12\,f^{-2}\dd f^2 -
\frac12\,f^{-2}(\dd \chi + v\dd u - u\dd v)^2 + f^{-1}\e^{-2\phi}\dd
v^2- f^{-1}\e^{2\phi}(\dd u - \kappa \dd v)^2 + 2\dd \phi^2 -
\frac12\,\e^{4\phi}\dd \kappa^2\,.
 \ee
This space has the isometry group $G=Sp(4,\mathbb{R})$, the same as
its Lorentzian counterpart \cite{emda}, in which case
 \be\lb{tar1} \dd l_L^2
=\frac12\,f^{-2}\dd f^2 + \frac12\,f^{-2}(\dd \chi + v\dd u - u\dd
v)^2 - f^{-1}\e^{-2\phi}\dd v^2 - f^{-1}\e^{2\phi}(\dd u - \kappa
\dd v)^2 + 2\dd \phi^2 + \frac12\,\e^{4\phi}\dd \kappa^2\,.
 \ee
The Euclidean line element (\ref{tar4}) is derived from (\ref{tar1})
by the following complexification:
 \be\lb{compl}
v\to  \ii v \,,\quad \chi\to \ii \chi\,,\quad \kappa\to - \ii
\kappa\,.
  \ee
The metric (\ref{tar4}) is the metric on the coset $G/H$, whose
nature can be uncovered from a signature argument. The Killing
metric of $sp(4,\mathbb{R})\sim so(3,2)$ algebra has the signature
$(+6,-4)$, with plus standing for non-compact and minus for compact
generators. Since the signature of the target space is $(+3,-3)$, it
is clear that the isotropy subalgebra contains three non-compact and
one compact generators. Such a subalgebra of $so(3,2) \sim
sp(4,\mathbb{R})$ is ${\rm lie\,}(H) \sim so(2,1)\times so(1,1) \sim
gl(2,\mathbb{R})$. We therefore deal with the coset
$SO(3,2)/(SO(2,1)\times SO(1,1))=Sp(4,\mathbb{R})/GL(2,\mathbb{R})$.

In addition to the bulk action we have a number of surface terms
resulting from three-dimensional dualizations as well as from
dimensional reduction of the four-dimensional Gibbons-Hawking-axion
term. Collecting these together, and taking care of the rescaling of
the electric and magnetic potentials, we get:
  \be\lb{btot}
S_{\rm inst} = \textsuperscript{3}S_b =
\frac{\beta}{16\pi}\int\limits_{\pa\cE} (-2[k]\ast\dd \Psi +
\ast\dd\psi_0) + \frac{\beta}{16\pi}\int\limits_{\pa\cE'}\left([\ka
\e^{4\phi}\ast \dd\ka] + 2\sqrt2uF + (\chi+uv)\cF\right) \,.
 \ee
Note that the {\em on-shell} value of the action which we are
interested in for instantons is entirely given by the boundary term
$\textsuperscript{3}S_b$ since the bulk sigma-model action vanishes
by virtue of the contracted three-dimensional Einstein equations
 \be
\cR_{ij}=G_{AB}\pa_i X^A\pa_j X^B\,.
 \ee

Variation of the bulk action (\ref{acsig}) over $X^A$ gives the
equations of motion
 \be\lb{eqsigma}
\pa_i\left(\sqrt{h}h^{ij}G_{AB}\pa_j X^B\right)=
 \fr12\, G_{BC,A}\pa_i X^B\pa_j X^C h^{ij}\sqrt{h},
 \ee
which can be rewritten in a form explicitly covariant both with
respect to the three-space metric $h_{ij}$, and to the target space
metric $G_{AB}$
 \be\lb{consJ}
\nabla_i J^i_{A}=0\,,
 \ee
where $\nabla_i$ is the total covariant derivative involving
Christoffel symbols  both of  $h_{ij}$ and $G_{AB}$. The six
currents associated with the potentials read
 \be
J^i_{A}=h^{ij}\pa_j X^B G_{AB}\,.
 \ee
Note that, according to~\eqref{tar4}, the NUT potential $\chi$ is a
cyclic target space coordinate, so the corresponding current
satisfies the conservation equation with an odinary derivative
 \be
\pa_i\left(\sqrt{h}h^{ij}J_{j\chi}\right)=0\,,\quad
J_{i\chi}=G_{\chi A}\pa_i X^A\,,
 \ee
and defines a conserved quantity, the NUT charge
 \be
N=-\int\limits_{\pa\cE}\sqrt{\si}n^iJ_{i\chi}\dd ^2 x=
 \fr12\int\limits_{\pa\cE}\sqrt{\si}f^{-2}n^i(\pa_i\chi+v\pa_i u- u\pa_i
 v)\dd ^2x.
 \ee
where $\pa\cE$ is any topological two-sphere and $n^i$ is the
outward normal.

\setcounter{equation}{0}
\section{Matrix representation}

To proceed, we have to introduce the matrix representation of the
coset $Sp(4,\mathbb{R})/GL(2,\R)$. The symplectic group
$Sp(4,\mathbb{R})$ is the group of real $4\times 4$ matrices $M$
satisfying
 \be
M^TJM = J\,,\quad J = \begin{bmatrix} 0 & \sigma_0 \\
-\sigma_0 & 0 \end{bmatrix}\,,
 \ee
where $\sigma_0$ is the $2\times 2$ identity matrix. The group
$Sp(4,\mathbb{R})$ has three maximal subgroups \cite{Gilmore} of
dimension four, the compact subgroup $U(2)$, and the two non-compact
subgroups $U(1,1)$ and $GL(2,\mathbb{R})$, leading to the three
cosets associated with three-dimensional reductions of EMDA:
$Sp(4,\mathbb{R}) /U(2)= SO(3,2)/(SO(3)\times SO(2))$ (reduction of
Lorentzian EMDA relative to a spacelike Killing vector, or of
phantom Lorentzian  EMDA relative to a timelike Killing vector
\cite{phantom2}), $Sp(4,\mathbb{R})/U(1,1) = SO(3,2)/(SO(2,1)\times
SO(2))$ (reduction of Lorentzian EMDA relative to a timelike Killing
vector), and $Sp(4,\mathbb{R})/GL(2,\mathbb{R}) =
SO(3,2)/(SO(2,1)\times SO(1,1))$ (reduction of normal or phantom
Euclidean EMDA).

We will use the representation
\cite{diak} for the $sp(4,\mathbb{R})$ algebra
\begin{eqnarray}
V_a=\frac{1}{2}\left(\begin{array}{cc}
0&\sigma_a\\
\sigma_a &0
\end{array}\right)\,,\;\; \;\; W_a=\frac{1}{2}
\left(\begin{array}{cc}
\sigma_a & 0\\
0&-\sigma_a
\end{array}\right)\label{p2}\,,\lb{nc}\\
U_a=\frac{1}{2} \left(\begin{array}{cc}
0&\sigma_a\\
-\sigma_a &0
\end{array}\right)\,,\;\; \;\; U_2=\frac{1}{2}
\left(\begin{array}{cc}
\sigma_2 &0\\
0&\sigma_2
\end{array}\right)\,,\lb{c}
\end{eqnarray}
($a = 0,1,3$, $\sigma_1 = \sigma_x$,
$\sigma_2 = i\sigma_y$, $\sigma_3 = \sigma_z$ with
$\si_x,\,\si_y,\,\si_z$ the Pauli matrices). The matrices $V_a,\,W_a$
are symmetric and the four matrices $U_a,\,U_2$ are antisymmetric.
An extensive discussion of the internal algebraic structures in this
matrix space can be found in \cite{bps}.

The isotropy subgroup $H = GL(2,R)$ for Euclidean EMDA leaves
invariant a given fixed point $\textbf{X}$ of the target space. It
is convenient to choose this point to be the point at infinity
$\textbf{X}( \infty)$, which will depend on the boundary conditions.
We shall assume that the three-space $\cE$ is asymptotically flat
and topologically $\mathbb{R}^3$, so that the asymptotic three-metric
is, in spherical coordinates,
 \be\lb{sph}
\dd\sigma^2 \equiv h_{ij}\dd x^i\dd x^j \simeq \dd r^2 +
r^2(\dd \theta^2 + \sin^2\theta \dd\varphi^2)\,.
 \ee
The possible asymptotic behaviors for $r = |\rr|\to\infty$ of the
three-dimensional fields $\textbf{X}(\rr)$ can in principle be
derived from the analysis of the three-dimensional field equations
or, equivalently, by a discussion of the possible algebraic types of
the matrix representing a given point of target space. In the
generic ALF case, the asymptotic four-dimensional metric is
 \be\lb{metalf}
\dd s^2 \simeq (\dd t - 2N\cos\theta\,\dd\varphi)^2 + \dd r^2 + r^2
(\dd\theta^2 + \sin^2\theta\,\dd\varphi^2)\,,
 \ee
and $f(\infty) = 1$, while the five other target space coordinates go to
zero
 \be\lb{alf}
\textbf{X}( \infty) = (1,0,0,0,0,0)\,.
 \ee
We will start with this asymptotic behavior as basis to build our matrix
representation of
$Sp(4,\mathbb{R})/GL(2,\mathbb{R})$, then discuss the non-ALF cases,
which are connected to the ALF case by group transformations.

In our representation of  $sp(4,\mathbb{R})$ the generators
(\ref{nc}) are non-compact, while (\ref{c}) are compact. We can
choose the $GL(2,\mathbb{R})$ subalgebra to be spanned by
 \be \lb{eiso}
{\rm lie\,}(H)= (V_a, U_2)\,,
 \ee
while ($U_a, W_a$) will be  the generators of the coset. The
infinitesimal transformations generated by (\ref{eiso}) leave
invariant the real $4\times4$ matrix
 \be \lb{etaE} \eta =
\left(\begin{array}{cc} \sigma_0 & 0 \\
0  & -\sigma_0
\end{array}\right)\,.
 \ee
A symmetric matrix representative $M$ of the coset such that
$M(\infty) = \eta$ for the ALF asymptotic behavior (\ref{alf}) has
the block structure
 \be\lb{ME} M =
\left(\begin{array}{cc}
P ^{-1}&P ^{-1}Q\\
QP ^{-1}&-P +QP ^{-1}Q
\end{array}\right)\,,
 \ee
with the same  $Q$ as in \cite{diak}, but a new $P$, namely
 \be\lb{PQ1} P = \e^{-2\phi}\left(\begin{array}{cc} f\e^{2\phi}+v^2&
v\\v&1
\end{array}\right)\,, \quad
Q=\left(\begin{array}{cc} vw -\chi & w\\ w &-\kappa
\end{array}\right) \quad (w=u-\kappa v)\,.
 \ee
The $2\times2$ block matrices in (\ref{ME}) are given by
 \ba\lb{block}
&&P^{-1} = \left(\begin{array}{cc} f^{-1} & -f^{-1}v\\
-f^{-1}v&\e^{2\phi}+f^{-1}v^2
\end{array}\right)\,,\quad P^{-1}Q= \left(\begin{array}{cc}
-f^{-1}\chi & f^{-1}u\\ f^{-1}v\chi+w\e^{2\phi} & -\kappa
\e^{2\phi}-f^{-1}vu
\end{array}\right)\,, \nn\\
&&-P + QP^{-1}Q = \left(\begin{array}{cc} -f - v^2\e^{-2\phi} +
w^2\e^{2\phi} + f^{-1}\chi^2 & -v\e^{-2\phi} - \kappa w\e^{2\phi}
-f^{-1}u\chi\\
-v\e^{-2\phi} - \kappa w\e^{2\phi} -f^{-1}u\chi& -\e^{-2\phi} +
\kappa^2 \e^{2\phi}+f^{-1}u^2 \end{array}\right)\,.
 \ea
We note that the matrix (\ref{ME}) is not symplectic, but
antisymplectic:
 \be
 M^TJM  = -J\,.
 \ee
However this is enough to ensure that the matrix current
 \be
J^i = h^{ij}M^{-1}\partial_jM
 \ee
is symplectic. The matrix (\ref{ME}) can be obtained from the corresponding
matrix in \cite{emda} as follows. Analytical continuation
(\ref{compl}), together with multiplication by $\ii$ of the second
row and column of the $2\times2$ blocks $P$ and $Q$ leads to the
blocks $P $ and $\ii Q$. Then multiplication by $-\ii$ of the second
row and column of the matrix $M$ in \cite{emda} leads to (\ref{ME}).
In terms of $M $ the target space metric (\ref{tar4}) will read
 \be\label{dl}
\dd l^2 = -\frac14\,\tr\left( \dd M \dd M ^{-1}\right) =
\frac12\,\tr\left[(P ^{-1}\dd P )^2 - (P ^{-1}\dd Q)^2\right]\,,
 \ee
while the sigma-model field equations (\ref{consJ}) read
 \be\lb{sigeq}
\pmb\nabla\left(M^{-1}\pmb\nabla M\right)=0\,,
 \ee
where $\pmb\nabla$ stands for the three--dimensional covariant
derivative, and the scalar product with respect to the metric $h_{ij}$
is understood.

If the ALF restriction (\ref{alf}) is raised, the representative
matrix (\ref{ME}) will go at infinity to an arbitrary constant
symmetric antisymplectic matrix
 \be
M(\infty) = A
 \ee
different from $\eta$. Generically this matrix will be of the form
(\ref{ME}) where the fields $f$, $\phi$, etc. are replaced by their
(arbitrary) values at infinity $f(\infty)$, $\phi(\infty)$, etc. As
the scalar potentials $\psi$, $\phi$, $v$, as well as the
pseudoscalar potentials $\chi$, $\kappa$, $u$, are only defined up
to an additive constant, the generic $M(\infty)$ can always be
gauge-transformed to the ALF form $\eta$. An exceptional
$M(\infty)$, which is not gauge-equivalent to $\eta$, is one for
which (\ref{ME}) breaks down (at infinity) because $P^{-1}(\infty)$
is not invertible, i.e. det$(P^{-1})(\infty) \equiv
(f^{-1}\e^{2\phi})(\infty) = 0$. This can be subclassified according
to the rank of $P^{-1}(\infty)$. Rank 1 corresponds to either
$f^{-1}(\infty) = 0$ or $\e^{2\phi}(\infty) = 0$, while rank 0
($P^{-1}(\infty) = 0$) corresponds to both vanishing. Let us discuss
briefly these three possible exceptional asymptotic behaviors (more
details are given in Appendix B):

{\em Case E1} (ALE). In the case $f^{-1}(\infty) = 0$, the
asymptotic solution of the sigma model field equations
(\ref{sigeq}), which is
 \be
M(r) \simeq A(I + Br^{-1})
 \ee
(with $B$ a constant symplectic matrix) leads to $f^{-1} = M_{11}
\simeq O(r^{-1})$, which may be normalized to
 \be\lb{fr}
f \simeq r\,.
 \ee
As shown in Appendix B, the asymptotic coset representative $A$
can be gauge transformed to $A=\eta'_1$ with
 \be
\eta'_1 = \left(\begin{array}{cccc} 0 & 0 & \pm1 & 0 \\
0 & 1 & 0 & 0 \\  \pm1 & 0 & 0 & 0 \\ 0 & 0 & 0 & -1
\end{array}\right) \,.
 \ee
It follows that $\chi = -M_{13}f$ goes asymptotically to $\chi
\simeq \mp f$ which, with asymptotically vanishing electromagnetic
potentials $v$ and $u$, is dualized using (\ref{twist}) to
$\omega_{\varphi} \simeq \pm\cos\theta$. The resulting asymptotic
four-dimensional metric (\ref{an}) is recognized to be the
four-dimensional Euclidean metric in three-spherical coordinates
 \be\lb{4eucl}
ds^2 \simeq \dd \rho^2 + \rho^2 \dd \Omega_3^2 = \dd\rho^2 +
\frac{\rho^2}4[\dd \theta^2 + \sin^2\theta \dd \varphi^2 + (\dd \eta
\mp \cos\theta \dd \varphi)^2]\,,
 \ee
with the angular coordinate $\eta = t$, and the radial coordinate
$\rho = (4r)^{1/2}$. This is the ALE case.

{\em Case E2}. In the case $\e^{2\phi}(\infty) = 0$ with
$f^{-1}(\infty) \neq 0$, the asymptotic matrix representative can be
gauge transformed to
 \be
\eta'_2 = \left(\begin{array}{cccc} 1 & 0 & 0 & 0 \\
0 & 0 & 0 & \mp1 \\  0 & 0 & -1 & 0 \\ 0 & \mp1 & 0 & 0
\end{array}\right)\,.
 \ee
This is an exceptional ALF case.

{\em Case E3}. In the case $P^{-1}(\infty) = 0$, the asymptotic
matrix representative can be gauge transformed to the block form
 \be
\eta'_3 = \left(\begin{array}{cc} 0 & \beta \\ \beta & 0
\end{array}\right)\,, \quad \beta^2 = 1\,,
 \ee
as shown in Appendix B. This includes an exceptional ALE subcase
E3a, with
 \be
\beta_{\rm a} = \left(\begin{array}{cc} \pm1 & 0 \\ 0 & \pm1
\end{array}\right)\,,
 \ee
and a one-parameter subcase E3b, with
 \be
\beta_{\rm b} = \left(\begin{array}{cc} \cos\nu & \sin\nu \\ \sin\nu &
-\cos\nu
\end{array}\right)\,,
 \ee
interpolating between a second ALE behavior for $\sin\nu = 0$, and a
magnetic linear dilaton asymptotic behavior with linearly rising
gravitational, dilaton and magnetic potentials (the magnetic
Euclidean equivalent of the electric linear dilaton behavior in
Lorentzian EMDA \cite{Clement:2002mb}) for $\cos\nu = 0$.

\setcounter{equation}{0}
\section{Null geodesic solutions}

In the following  we will use the formalism developed in
\cite{diak,bps}. For the reader's convenience we reproduce here
basic results. Starting with the sigma-model action in the matrix
form
\begin{equation}\label{ngs}
S_\sigma= -\frac{\beta}{16\pi}\int \dd ^3x\sqrt{h}\left\{{\cal R}+
\frac{1}{4}\,\tr ( \pmb\nabla M \pmb\nabla M^{-1})\right\}\,\,,
\end{equation}
we obtain the equations of motion (\ref{sigeq}) together with the
three--dimensional Einstein equations
\begin{equation} \label{ei}
{\cal R}_{ij}=-\frac{1}{4}\,\tr\left(\nabla_i M \nabla_j
M^{-1}\right)\,.
\end{equation}

As was noticed by Neugebauer and Kramer \cite{nk}, when one makes
the special assumption that all target space coordinates $X^A$
depend on $x^i$ through only one scalar potential, i.e.
$X^A=X^A[\tau(x^i)]$, it follows from the equation of motion  that
this potential can be chosen to be harmonic\footnote{\label{foo}Note
that if $\tau$ is harmonic in three dimensions ($\Delta_3 \tau=0$),
it will be harmonic in four dimensions
 ($\Delta_4 \tau=0$) as well, since
$\Delta_4={\sqrt{g}}^{-1}\pa_\mu \sqrt{g}
g^{\mu\nu}\pa_\nu=f{\sqrt{h}}^{-1}\pa_i \sqrt{h}
h^{ij}\pa_j=f\Delta_3$ \,.},
\begin{equation}
\Delta\tau=0\,, \qquad \Delta={\pmb\nabla}^2\,,
\end{equation}
Eq. (\ref{eqsigma}) reducing then to the geodesic equation on the
target space
\begin{equation}
\frac{\dd ^2X^A}{\dd \tau^2}+\Gamma^A_{BC}\,\frac{\dd X^B}{\dd \tau}
\,\frac{\dd X^C}{\dd \tau}=0\,.
\end{equation}
This may be rewritten in matrix terms as
\begin{equation}
\frac{\dd}{\dd \tau}\left(M^{-1}\,\frac{\dd M}{\dd \tau}\right)=0\,,
\end{equation}
and first integrated by
 \be\lb{B}
M^{-1}\frac{\dd M}{\dd\tau} = B \,,
 \ee
where $B\in {\rm lie\,}(G)\ominus{\rm lie\,}(H)$ is a constant matrix.
A second integration leads to the solution to the geodesic equation
in the exponential form
\begin{equation} \label{AB}
M = A\,{\rm e}^{B\tau}\,,
\end{equation}
with $A \in G/H$ another constant matrix. The parametrisation
(\ref{AB}) reduces the three--dimensional Einstein equations
(\ref{ei}) to
\begin{equation}
{\cal R}_{ij}=\frac{1}{4}\,(\tr B^2)\nabla_i \tau \nabla_j \tau\,.
\end{equation}
\noindent From this expression it is clear that in the particular
case
\begin{equation} \label{null}
\tr B^2 =0
\end{equation}
the three--space is Ricci--flat. In three dimensions the Riemann
tensor is then also zero, and consequently the three--space $\cE$ is
flat. We shall assume in the following that $\cE =\mathbb{R}^3$.
From Eq. (\ref{dl}) one can see that the condition (\ref{null})
corresponds to null geodesics \cite{gc} of the target space
\begin{equation}\lb{ngeo}
\dd l^2=\frac{1}{4}\,(\tr B^2)\,\dd \tau^2=0\,.
\end{equation}

An important feature of the target space of Euclidean EMDA
(\ref{tar4}) as compared to that of the Lorentzian theory
(\ref{tar1}) is that it has now the signature $(+,+,+,-,-,-)$ with
three, rather than two, independent null directions. Each null
direction gives rise to some BPS solution which is potentially
supersymmetric within a suitable supergravity embedding. One new
minus sign is associated with the twist potential, reflecting the
possibility of extremal Taub-NUT solutions (and consequently
multi-Taub-NUTs). Another new minus sign is related to the axion
field, reflecting the possibility of extremal instantons without
Maxwell fields. At the same time, the electric direction has now a
positive definite metric component, while the magnetic one remains
negative. So, in absence of twist and axion field, only magnetic or
dyonic configurations can be extremal.

Our boundary conditions imply that the harmonic potential
$\tau(x^i)$ goes to a constant value at infinity, which we can take
to be zero by a redefinition of the matrix $A$. Then, these
solutions are null target space geodesics going through the point
$A=M(\infty)$. In the following we discuss the ALF case, with
$A=\eta$,
 \be\lb{ngalf}
M = \eta\,\e^{B\tau}\,.
 \ee
Null geodesic going through other points $A = \eta'$ corresponding
to exceptional asymptotics,
 \be
M' = \eta'\,\e^{B'\tau}\,,
 \ee
can be generated from (\ref{ngalf}) by $Sp(4,\mathbb{R})$
transformations
 \be\lb{transf}
M' = K^TMK\,, \quad B' = K^{-1}BK\,.
 \ee

In the ALF case, the generators of the coset are $W_a,\,U_a$, so
that one can write
\begin{equation} \label{AB1}
B = 2(\alpha^aW_a+\beta^aU_a)\,,
\end{equation}
where $\alpha^a,\;\beta^a$ are constants depending on the charges.
These charges are the mass $M$ and NUT charge $N$, the dilaton and
axion charges $D,\, A$ and the electric and magnetic charges $Q,\,P$
defined, as in the Lorentzian case \cite{bps}, by the following
behavior of the target space variables at spatial infinity:
 \begin{eqnarray} \label{as}
f  \sim  1-\frac{2M}{r}\,,  \quad &&\chi  \sim    -\frac{2N}{r}\,,\nonumber\\
\phi   \sim  \frac{D}{r}\,,\quad &&  \kappa \sim \frac{2A}{r}\,, \nonumber\\
v \sim   \frac{\sqrt{2}Q}{r}\,, \quad && u\sim\frac{\sqrt{2}P}{r}\,.
\end{eqnarray} Using the
representation (\ref{nc}), (\ref{c}) for the generators we obtain
$B$ in the following block form: \be\label{e3} B=
\left(\begin{array}{cc}
    a&b\\
    -b&-a
\end{array}\right) \,,\quad  a=\alpha^a\sigma_a\,,\;
 b=\beta^a\sigma_a\,\;(a=0,1,3)\,,
 \ee
with symmetric $2\times 2$ blocks $a,\;b$. Assuming that the
monopole harmonic function is normalized to $\tau = r^{-1}$, and
comparing with (\ref{as}), we can express the coefficients and the
matrices in~\eqref{e3} in terms of the charges:
 \be\lb{albt}
\al^a=(M+D,\;-\sqrt{2} Q,\; M-D)\,,\quad \bt^a=(N-A,\;\sqrt{2} P,\;
N+A)\,,
 \ee
\begin{equation}\label{mab}
  a = \begin{pmatrix}
  2M & -\sqrt{2}Q\\
  -\sqrt{2}Q & 2D
  \end{pmatrix}\,,\quad
  b = \begin{pmatrix}
  2N & \sqrt{2}P\\
  \sqrt{2}P & -2A
  \end{pmatrix}\,.
\end{equation}
Note that the dualized one-forms $\dd\kappa$, $g$ and $\varpi$ may
be extracted from the lower left-hand block of $M^{-1}\dd M = B\dd\tau$:
 \be\lb{dualcon}
(M^{-1}\dd M)_{21} = -P^{-1}\dd QP^{-1} = \begin{pmatrix}
\varpi & - g - v\varpi\\ - g - v\varpi & \e^{4\phi}\dd\kappa + 2vg
+ v^2\varpi \end{pmatrix} = -b\dd\tau\,.
 \ee
In particular, $\varpi = -2N\dd\tau$, leading after inverse dualization
to
 \be\lb{nut}
\omega = \ast\varpi = 2N\cos\theta\dd\varphi
 \ee
for all monopole ALF geodesic solutions.

The matrix $B$ is identically traceless:
 \be\lb{btr}
 \tr B\equiv 0\,,
  \ee
while for $B^2$ one obtains
 \be\lb{b2tr}
\tr B^2=4\left[(\al^a)^2-(\bt^a)^2\right],
  \ee
where the Euclidean norm is understood:
$(\al^a)^2=(\al^0)^2+(\al^1)^2+(\al^3)^2$. Therefore the null
geodesic condition $\tr B^2=0$ translates into $
(\al^a)^2=(\bt^a)^2$, or in terms of charges
\begin{equation}\label{BPS}
    M^2+D^2+Q^2=N^2+A^2+P^2\,.
\end{equation}
This no-force condition for instantons can be obtained from that in
the Lorentzian sector \cite{bps}
\begin{equation}\label{BPSl}
    M^2+N^2+D^2+A^2=Q^2+P^2\,.
\end{equation}
by the complexification
 \be\lb{compl1} Q\to  \ii Q \,,\quad N\to \ii
N\,,\quad A\to - \ii A\,.
 \ee
corresponding to (\ref{compl}).

In the space of charges the group $H=SO(2,1)\times SO(1,1)$ is
operating as a duality symmetry, so it is convenient to replace the
Euclidean vectors $\al^a,\, \bt^a$ by the $SO(2,1)$ vectors
\begin{equation}\label{e4}
    \pmb{\mu} = (\mu^0,\overrightarrow{\mu})\equiv
    (\beta^0,\overrightarrow{\alpha})\,,\quad \pmb{\nu} =
    (\nu^0,\overrightarrow{\nu})\equiv (\alpha^0,\overrightarrow{\beta})\,,
\end{equation}
with $\overrightarrow{\alpha}\equiv (\alpha^1,\alpha^3)$ (similarly
for other variables). In terms of the charges,
\begin{equation}\label{mun}
    \pmb{\mu} = (N-A,\;-\sqrt{2} Q,\; M-D),\quad
    \pmb{\nu}=(M+D,\;\sqrt{2} P,\;N+A)\,.
\end{equation}
With this new parametrization, \eqref{AB1} takes the form
\begin{equation}\label{form}
    B(\pmb\mu,\pmb\nu) =
    2(\mu^0U_0+\mu^1W_1+\mu^3W_3+\nu^0W_0+\nu^1U_1+\nu^3U_3)\,.
\end{equation}
The condition (\ref{BPS}) now reads
\begin{equation}\label{e5}
    \pmb{\mu}^2 = \pmb{\nu}^2\,,
\end{equation}
with the $SO(2,1)$ norm: \be
\pmb{\mu}^2=\eta^{ab}\mu_a\mu_b\,,\;\;\;\eta^{ab}=\rm{diag}(-1,1,1)\,.
\ee This leads to
\begin{equation}\label{e6}
 B^2 =  2\left(\begin{array}{cc}
    \la^1\si_3-\la^3\si_1 & \la^0\sigma_2\\
   \la^0\sigma_2 & \la^1\si_3-\la^3\si_1
\end{array}\right) \,,
\end{equation}
where $\pmb{\la}$ is the skew product: \be\label{s1} \pmb{\la}=
\pmb{\mu}\wedge \pmb{\nu}\,,\quad \la_a
=\epsilon_{abc}\mu^b\nu^c\,,\quad\epsilon_{013}=+1\,. \ee For the
matrix $B^3$ one finds
 \be\lb{b3b}
B^3 = 2B[\n\wedge\pmb{\la},\m\wedge\pmb{\la}]\,,
 \ee
leading to
 \be\lb{b3tr}
\tr B^3=0\,.
 \ee
In view of (\ref{btr}), (\ref{null}) and (\ref{b3tr}), the
characteristic equation for $B$ reduces to
 \be\label{s2}
B^4+(\det B)I=0\,,
 \ee
so that $B^4$ is proportional to the unit $4\times 4$ matrix:
\begin{equation}\label{e7}
 B^4 =  4\pmb{\la}^2 I \,.
\end{equation}
In terms of the charges, using~\eqref{e5},
 \be\lb{la2ch}
\pmb{\la}^2 =[\pmb{\mu}\cdot\pmb{\nu}-\pmb{\mu}^2]
[\pmb{\mu}\cdot\pmb{\nu}+\pmb{\mu}^2]=
-[2m_+d_--q_-^2][2m_-d_+-q_+^2]\,,
 \ee
with
 \be
m_{\pm} = M \pm N\,, \quad d_{\pm} = D \pm A\,, \quad q_{\pm} = Q
\pm P\,.
 \ee
Note that the algebraic properties (\ref{btr}), (\ref{null}),
(\ref{b3tr}) and (\ref{s2}) of the matrix $B$, which have been
established in the ALF case, are also valid in the case of
exceptional asymptotic behaviors, the corresponding $B$ matrices
being related to those of the ALF case by the similarity
transformations (\ref{transf}).

Finally we evaluate the boundary action (\ref{btot}) for
null-geodesic solutions. This is the sum $S_{\rm inst}= S_1 + S_2$
of two surface integrals. The first, purely gravitational
contribution (\ref{bG}), is the sum of the regularized
Gibbons-Hawking term (\ref{redGH}) and the boundary integral
(\ref{b1}) for the background solution, both evaluated on a large
sphere of radius $R$. For an
ALF metric of the form (\ref{an}) with $\omega_i \dd x^i =
-2N\cos\theta \dd\varphi$, the appropriate background
\cite{Hunter:1998qe} is self-dual Taub-NUT
 \be
\dd s_0^2 = f_0(r_0)(\dd t_0 - 2N_0\cos\theta\,\dd\varphi)^2 +
f_0^{-1}(r_0)[\dd r_0^2 + r_0^2 (\dd\theta^2 +
\sin^2\theta\,\dd\varphi^2)]\,,
 \ee
with $f_0(r_0) = (1+2|N_0|/r_0)^{-1}$, and the matching conditions
are $t_0 = m(R)t$, $r_0 = m^{-1}(R)r$, $N_0 = m(R)N$, with $m(R)=
(f(R)/f_0(m^{-1}R))^{1/2} = 1 + (M-|N|)/R + O(R^{-2})$. The
regularized trace of the extrinsic curvature of $\pa\cM$ is, in the
monopole case,
 \be\lb{regKk}
[K] = K - K_0 = f^{1/2}(r)\left(k(r)-\frac12f^{-1}(r)f'(r)\right) -
f_0^{1/2}(r_0)\left(k_0(r_0)-\frac12f_0^{-1}(r_0)f'(r_0)\right)\,,
 \ee
with the extrinsic curvatures of $\pa\cE$ for the solution and the
background
 \be
k(r) = \frac2r\,, \quad k_0(r_0) = \frac2{r_0} =
\left(\frac{f(R)}{f_0(m^{-1}R)}\right)^{1/2}\frac2r\,.
 \ee
The net regularized extrinsic curvature $[k]$ of $\pa\cE$ ($r=R$) is
thus zero, so that (\ref{bG}) reduces to
 \be\lb{S1}
S_1 = -\frac{\beta}{16\pi}\lim_{R\to\infty}\int_{r=R}\sqrt{h}\dd \sigma
f^{-1/2}f_0^{-1/2}f'_0 = - \frac{\beta|N|}2\,.
 \ee

The second surface integral is that of the contributions of the
various dualizations evaluated on the boundary $\pa\cE'$, which has
two disjoint components, a large sphere at infinity with the normal
oriented outwards, and a small sphere shielding the source $r=0$ of
the harmonic potential $\tau = 1/r$ (we will later generalize to the
case of multi-center harmonic potentials) with the normal oriented
inwards:
 \ba\lb{Sb}
S_{2} &=& \frac{\beta}{16\pi}\oint_{\pa\cE'}\sqrt{h}\dd \sigma [
\e^{4\phi}\kappa\kappa'_r + f^{-2}(\chi+uv)(\chi'_r+vu'_r-uv'_r) +
2f^{-1}\e^{2\phi}u(u'_r-\kappa v'_r)] \nn\\
&=& \frac{\beta}4\left[\e^{4\phi}\kappa\dot\kappa + f^{-2}(\chi+uv)
(\dot\chi+ v\dot{u} - u\dot{v}) + 2f^{-1}\e^{2\phi}u(\dot{u}-\kappa
\dot{v}) \right]_{\tau=0}^{\tau=\infty}\,,
 \ea
where $\dot{}$ is the derivation relative to $\tau$. This may be
evaluated using the first integral (\ref{B}). The upper left-hand
corner block of (\ref{B}) gives
 \be
-\dot{P}P^{-1} + QP^{-1}\dot{Q}P^{-1} = B_{11} \equiv a \,.
 \ee
Tracing the different terms yields
 \ba
 -\tr (\dot{P}P^{-1}) &=& \frac{\dot\Delta}{\Delta} \qquad (\Delta \equiv
 \det(P^{-1}) = f^{-1}\e^{2\phi})\,, \lb{del}\\
 \tr (QP^{-1}\dot{Q}P^{-1}) &=& \e^{4\phi}\kappa\dot\kappa + f^{-2}(\chi+uv)
 (\dot\chi + v\dot{u} - u\dot{v}) +  2f^{-1}\e^{2\phi}u(\dot{u}-\kappa \dot{v})\,.
 \ea
This last term is the integrand of (\ref{Sb}). So,
 \be\lb{S2}
S_{2} = \frac{\beta}4\left[\tr (a) -
\frac{\dot\Delta}{\Delta}\right]_{\tau=0}^{\tau=\infty} =
-\frac{\beta}4\left[\frac{\dot\Delta}{\Delta}\right]_{\tau=0}^{\tau=\infty}\,.
 \ee
Summing (\ref{S1}) and (\ref{S2}) leads to the total action
 \be\lb{acdel}
S_{\rm inst} = \frac{\beta}4\left(
-\left[\frac{\dot\Delta}{\Delta}\right]_{\tau=0}^{\tau=\infty}-2|N|\right)\,.
 \ee

This shall be evaluated later in the various cases.

\setcounter{equation}{0}
\section{Discussion of the solutions: ALF asymptotics}

The generic matrix $B$ satisfying Eq. (\ref{s2}) is regular
(non-degenerate) and of rank 4. If $B$ is singular (degenerate),
det$\,B=0$, the exponential in (\ref{ngalf}) reduces to a polynomial
of third degree if rank$\,B=3$ (we will qualify this case as
``weakly degenerate'') or, as we shall see, of first order if
rank$\,B=2$, which will be the ``strongly degenerate'' case. In this
section we will investigate these three classes of null geodesic
solutions in the case of ALF asymptotics, and treat the case of
exceptional asymptotics in the next section.

\subsection{Strongly degenerate case}

According to (\ref{e7}) $B$ is degenerate if $\pmb{\la}$ is
lightlike,
 \be\lb{deg}
\pmb{\la}^2 = 0\,.
 \ee
On account of (\ref{e5}), this corresponds to the condition on the
charge vectors,
 \be\lb{dege}
(\pmb{\mu}+\epsilon\pmb{\nu})^2=0\,,\quad \epsilon =\pm 1\,,
 \ee
or, in terms of the charges
\begin{equation}\label{fc}
2m_{\epsilon}d_{-\epsilon} - q_{-\epsilon}^2 = 0\,.
\end{equation}
Contrary to the Lorentzian case \cite{bps}, the vanishing
(\ref{deg}) of the square of the $SO(2,1)$ vector $\pmb{\la}$ does
not imply the vanishing of this vector itself. Thus, Eq~\eqref{e5}
along with the generic condition~\eqref{dege} lead to the weakly
degenerate case ${\rm rank}\,B=3$. However, if the stronger
condition
 \be\lb{laze}
\pmb{\la}=0
 \ee
is satisfied along with~\eqref{e5}, then ${\rm rank}\,B=2$ and
$B^2=0$. In this strongly degenerate case, the matrix $M$ depends
linearly on $\tau$
 \be \lb{ML}
M=\eta(I+B\tau)\,.
 \ee
Comparing with (\ref{ME}), one obtains
 \ba\lb{potlin}
&& f = (1+2M\tau)^{-1} \,,\quad \chi=-2Nf\tau =
 \frac{N}{M}(f-1)\,,\nn\\
&& \e^{2\phi} = 1+2D\tau-2Q^2f\tau^2\,,\quad \kappa =
2\e^{-2\phi}\tau(A-PQf\tau) \,,\\ && v=\sqrt{2}Qf\tau \,,\quad
u=\sqrt{2}Pf\tau \,. \nn
 \ea
In the special case of a one-center harmonic function $\tau=1/r$ the
resulting metric is, on account of (\ref{nut}),
 \be\lb{metlin}
\dd s^2=\left(1+\fr{2M}{r}\right)^{-1}(\dd t-2N\cos\theta \dd \varphi)^2+
\left(1+\fr{2M}{r}\right)\left[\dd r^2+r^2(\dd\theta^2 +
\sin^2\theta\dd\varphi^2)\right].
 \ee
This generically has non-zero Ricci tensor, with scalar curvature
and Kretchmann invariant
\begin{align}
& R=g^{\mu\nu}R_{\mu\nu}=\frac{2 (M^2-N^2)}{r (2 M+r)^3}\,,\\
& R^{\mu\nu\rho\si}R_{\mu\nu\rho\si}=\frac{44 (M^2-N^2)^2+64 M
(M^2-N^2)r +48 (M^2+N^2) r^2}{r^2 (2 M+r)^6}\,,
\end{align}
so that there is a curvature singularity at $r=0$, unless $N=\pm M$
(see below), in which case (\ref{metlin}) is a regular vacuum
metric, namely (anti-)self-dual Taub-NUT. We also note that the
result (\ref{potlin}) implies
 \be\lb{dilin}
\Delta = 1+2(M+D)\tau+2(2MD-Q^2)\tau^2\,,
 \ee
so that, depending on the values of the charges, this linear
solution may also develop a singularity for a finite value of
$\tau$. Excluding curvature singularities for a finite $r$ by the
constraint $M>0$, we find that (\ref{dilin}) inserted in (\ref{acdel})
always leads to a finite action
 \be\lb{acsd}
S_{\rm inst} = \frac{\beta}2(M-|N|+D)
 \ee
for a single center, and
 \be
S_{\rm inst} = \frac{n\beta}2(M-|N|+D)
 \ee
for a multi-center solution $\tau = \sum_{i=1}^n 1/|{\bf r}- {\bf
r}_i|$, irrespective of the possible presence of singularities of
$\e^{2\phi}$. In the vacuum case $D=A=Q=P=0$ and $N=\pm M$ from the
no-force condition, so that the one-center solution reduces to the
self-dual Taub-NUT instanton with vanishing action.

The strong degeneracy condition $\pmb{\la}=0$ holds if the two
vectors $\pmb{\mu}$ and $\pmb{\nu}$ are collinear, with either one
of the vectors vanishing as limiting cases. The generic condition
\be\lb{col} \pmb{\nu}=c\pmb{\mu}\,,\quad c=-P/Q\,, \ee splits into
two subcases:

1) If the vectors $\pmb{\mu}$ and $\pmb{\nu}$ are not necessarily
lightlike, one must have $c=-\epsilon$ in view of (\ref{dege}). This
implies
 \be\lb{charA1}
N=-\epsilon M\,, \quad A=\epsilon D\,,\quad P=\epsilon Q\,,
 \ee
so that only three of the charges are independent. These solutions,
where the no-force condition (\ref{BPS}) is solved by independently
balancing each electric-type charge by an equal magnetic-type
charge, generalize the Taub-NUT instantons of \cite{inst}. In the
case of a one-center harmonic function $\tau=1/r$, the corresponding
metric is the vacuum (anti-) self-dual Taub-NUT
 \be
\dd s^2=\left(1+\fr{2M}{r}\right)^{-1}(\dd t+2\eps M\cos\theta \dd
\varphi)^2+ \left(1+\fr{2M}{r}\right)\left[\dd r^2+r^2(\dd
\theta^2+\sin^2\theta \dd \varphi^2) \right]\,,
 \ee
where to remove the Misner string singularity one must identify $t$
with the period $8\pi M$. More generally, both the Maxwell field and
the axidilaton fields are separately self-dual, so that the
corresponding energy-momentum tensors vanish. The relations
(\ref{potlin}) and (\ref{charA1}) lead for the dilaton-axion system
to
 \be\lb{axidul}
\ka =\eps (1-\e^{-2\phi})\,,
 \ee
implying cancellation of the scalar terms at the right hand side of
the four-dimensional Einstein equations (\ref{einst4}), and for the
Maxwell system to
 \be
{\tilde F}_{i4}=\frac{1}{\sqrt{2}}\,\e^{2\phi}
(\kappa\partial_iv-\partial_iu) = -\frac{\eps}{\sqrt2}\,\partial_iv
= -\eps F_{i4}\,,
 \ee
leading to cancellation of the Maxwell terms. Therefore the subcase
1 strongly degenerate solution represents the self-dual EMDA
dressing of the Ricci-flat self-dual Taub-NUT instanton, with the
finite action
 \be\lb{acnd}
S_{\rm inst} = 4\pi |N|D
 \ee
(except in the case of a cylindrical spacetime, $|N|=M=0$, in which
case $S_{\rm inst} = \beta D/2$ with $\beta$ arbitrary). To our
knowledge, this non-vacuum instanton has not appeared in the
literature before (its Lorentzian counterpart, however, is known
\cite{kkot,bps}).

Actually, this subcase should be divided into three sectors,
according to the sign of the pseudonorm \be \pmb\mu^2 = \pmb\nu^2 =
2(Q^2-2MD)\,. \ee
\begin{description}
  \item[ ] 1a) Timelike sector ($Q^2<2MD$). All solutions of this
  sector can be generated by $SO(2,1)$ transformations from the
  neutral $\overrightarrow{\mu} = \overrightarrow{\nu} = 0$ solution with
  $P=Q=0$, $A=-N=\epsilon D = \epsilon M$. This sector can be further divided
  into future and past (for the vector $\pmb\nu$). In the future
  timelike sector ($M>0$, $D>0$), (\ref{potlin}) shows that the
  exponentiated dilaton $\e^{2\phi}$ and the metric function $f$ are
  obviously positive for all positive $\tau$, so that these
  solutions are regular for a multicenter harmonic function
   \be\lb{multi}
  \tau = \sum\limits_{i=1}^s\frac1{|\rr-\rr_i|}\,,
   \ee
with equal residues to ensure absence of Misner strings if $t$ is
periodically identified with period $8\pi M$. This is the EMDA
dressed generalisation of the multi-Taub-NUT instanton of Gibbons
and Hawking. In the past timelike sector ($M<0$, $D<0$), both
$\e^{2\phi}$ and $f$ develop a singularity for a finite positive
value of $\tau$.

  \item[ ] 1b) Lightlike sector ($Q^2=2MD$). This relation is
  reminiscent of a similar relation in the Lorentzian sector
  \cite{bps} $d=-q^2/2m$, with the complex charges $q=Q+\ii P$,
  $m=M+\ii N$, $d=D+\ii A$. So in some sense the solutions of this sector
  can be considered as analytic continuations of stationary extremal
  solutions to EMDA. Again, this sector can be divided in a
  future lightlike sector ($M$ and $D$ positive), with regular multi-Taub-NUT
  instantons as above, and a past lightlike sector ($M$ and $D$ negative) where
  $\e^{2\phi}$ and $f$ become singular for a finite positive value of $\tau$.

  \item[ ] 1c) Spacelike sector ($Q^2>2MD$). All the solutions of this
  sector, which can be generated by $SO(2,1)$ transformations from the
  neutral solution with $P=Q=0$, $A=N=\epsilon D = -\epsilon M$, lead to a
  singular $\e^{2\phi}$.
\end{description}

2) If the vectors $\pmb{\mu}$ and $\pmb{\nu}$ are lightlike
($\pmb{\mu}^2 = \pmb{\nu}^2 = 0$), then $c \neq \pm 1$ remains an
arbitrary parameter. In addition to (\ref{BPS}) one has two more
constraints on the charges
  \be\lb{col1}
  (D+M)Q=(A-N)P\,,\quad (D-M)P=(A+N)Q\,,
  \ee
so that again only three of the charges are independent (note that
the relations (\ref{charA1}) also solve the conditions
(\ref{col1})). Other relations between the charges (which follow
from the preceding) are
  \be
  D^2-A^2=M^2-N^2=(P^2-Q^2)/2\,,\quad PQ=AM-ND\,.
  \ee
  This subcase includes in the limits $c\to0$ and $c\to\infty$:
\begin{description}
  \item[ ] --- A 2-charge family of purely electric solutions if $\pmb{\nu}=0$,
with
  \be
  N^2=M^2 +Q^2/2\,,\quad D=-M\,,\quad A=-N\,,\quad  P=0\,.
  \ee
These solutions have negative action
 \be
S_{\rm inst} = -4\pi N^2
 \ee
(with $\beta=8\pi|N|$), which can be correlated with the fact that,
from Eq. (\ref{dilin}),
 \be
 \e^{2\phi} = f[1-4N^2\tau^2]\,,
 \ee
showing that they develop a singularity for a finite value of
$\tau$.
  \item[ ] --- A 2-charge family of purely magnetic solutions if
$\pmb{\mu}=0$, with
  \be\lb{2magn}
  N^2= M^2-P^2/2\,,\quad D=M\,,\quad A=N\,,\quad  Q=0\,.
  \ee
These solutions are regular for $f>0$ (but singular for $f=0$ if $P
\neq 0$), with \be \e^{-2\phi} = f, \quad v = 0, \quad \chi = \kappa
= -\sqrt{2}\frac{N}{P}u = \frac{N}{M}(f-1)\,. \ee
\end{description}

A special class in this subcase is that of neutral solutions with
$P=Q=0$. Then, the relations $\pmb{\mu}^2 = \pmb{\nu}^2 = 0$ and
$\pmb{\nu} = c\pmb{\mu}$ are solved by
 \be\lb{charA2n}
N=\epsilon' M\,, \quad A=\epsilon' D\,,\quad P=Q=0\,,
 \ee
with $\epsilon' = \pm1$. These relations (note the difference with
Eq. (\ref{charA1}) for the strongly degenerate subcase A1) lead to a
solution which is also a generalization of the Taub-NUT instanton,
again supporting a self-dual axidilaton,
 \be \e^{2\phi} = 1 +
2D\tau\,, \quad \ka =\eps' (1-\e^{-2\phi})\,.
 \ee
This solution is regular for positive $\tau$ provided both $M$ and
$D$ are positive, leading again to a positive action (\ref{acnd}).

\subsection{Weakly degenerate case}

This is the generic case $\pmb{\la}\neq 0, \;\pmb{\la}^2=0$
corresponding to ${\rm rank}\,B=3$ and $B^3\neq 0$, however $B^4=0$
since $\pmb{\la}^2=0$. The expression for $M$ includes three powers
of $\tau$
\begin{equation}\label{w1}
    M=\eta(I+B\tau+B^2\tau^2/2+B^3\tau^3/6)\,.
\end{equation}
Because of this cubic behavior of the matrix representative, the
evaluation of the action (\ref{Sb}) is delicate owing to the
occurence of infrared divergences in the individual factors, but
leads directly to a finite result when the form (\ref{acdel}) is
used. The function $\Delta$ in (\ref{del}) is then a polynomial of
maximum degree 6 which is dominated for $\tau\to\infty$ by its
leading term, $\Delta \sim O(\tau^p)$ ($p \le 6$), leading to
$\dot\Delta/\Delta \propto \tau^{-1}$ for $\tau\to\infty$. On the other
hand, from the ALF behaviors (\ref{as}), $\dot\Delta/\Delta = 2(M+D)$
for $\tau=0$, leading to the same finite value for the boundary
action
 \be
S_{\rm inst} = \fr{\beta}2 (M-|N|+D)
 \ee
as in the case of strongly degenerate ALF instantons.

Since the six charges are now related by the two
conditions~\eqref{BPS} and~\eqref{fc}, the target space coordinates
are generally given in terms of four independent charges. The
relations between the charges are generically nonlinear, except in
the following two subcases 1) and 2) where these relations
linearize, leading to solutions depending on only three charges. By
virtue of their orthogonality to the lighlike vector $\pmb{\la}$,
the vectors $\pmb{\mu}$ and $\pmb{\nu}$ are spacelike, so that all
weakly degenerate solutions can be generated by $SO(2,1)$
transformations from either representative 1) or 2).

1) The relations
 \be\lb{r1}
N=\epsilon' M,\,A=\epsilon' D,\,P=\epsilon Q\,,
 \ee
with $\epsilon' = \pm1$ independently of $\epsilon$, obviously solve
Eqs. (\ref{fc}) and (\ref{BPS}). These relations (note again the
difference with Eq. (\ref{charA1}) for the strongly degenerate
subcase A1)) generalize the relations (\ref{charA2n}) defining the
neutral solution of case A2) and again lead to a generalization of
the Taub-NUT instanton, with again the action (\ref{acnd}). The
vectors $\pmb\mu$ and $\pmb\nu$ are given by
\begin{equation}\label{w2}
    \pmb{\mu} = (\epsilon_1(M-D),\,-\sqrt{2} Q,\, M-D)\,,\;
    \pmb{\nu}=(M+D,\,-\epsilon_1 \epsilon_2 \sqrt{2}Q,\,\epsilon_1(M+D))\,,
\end{equation}
with $\epsilon_1 = \epsilon'$, $\epsilon_2 = -\epsilon\epsilon'$.
The target space coordinates read
\begin{align}
& f^{-1}=1+2 M \tau +4(1+\epsilon_2) D Q^2 \tau ^3/3\,,\nn\\
& \chi = \epsilon_1(f-1)\,,\nn\\
& \e^{2\phi}=1+2 D \tau +4 (1-\epsilon_2 ) M Q^2 \tau ^3/3  - f^{-1}v^2\,,\nn\\
& \kappa = \epsilon_1\{1 - [1 - (1+\epsilon_2)\sqrt{2}Qv\tau]\e^{-2\phi}\}\,,\label{solw1}\\
& v=\sqrt{2}Qf\tau[1+(1+\epsilon_2 )D\tau + (1-\epsilon_2 )M \tau ]\,,\nn\\
& u=\epsilon_1[v- (1+\epsilon_2)\sqrt{2}Qf\tau]\,.\nn
\end{align}

In the case $\epsilon_2=-1$ ($\epsilon' = \epsilon$), the relations
(\ref{solw1}) simplify to
 \ba\lb{solw1e2-}
f &=& 1 + \epsilon_1\chi =
(1+2M\tau)^{-1}, \quad v = \epsilon_1u = \sqrt{2}Q\tau\,,\nn\\
\e^{2\phi} &=& 1 + 2D\tau - 2Q^2\tau^2 - \frac43MQ^2\tau^3\,, \quad
\kappa = \epsilon_1[1 - \e^{-2\phi}]\,,
 \ea
so that again the Maxwell and axidilaton fields are separately
self-dual, leading for $M > 0$ to a regular metric which is that of
the Taub-NUT instanton. However, the associated dilaton becomes
singular at a finite distance from the centers $\tau \to \infty$.

In the case $\epsilon_2=+1$ ($\epsilon' = -\epsilon$), the metric
is, on account of (\ref{nut}),
 \be\lb{worm}
\dd s^2= f(r)(\dd t-2N\cos\theta \dd \varphi)^2+ f^{-1}(r)\left[\dd
r^2+r^2(\dd \theta^2+\sin^2\theta \dd \varphi^2) \right]\,,
 \ee
with
 \be
f^{-1}(r) = 1+\fr{2M}{r}+\fr{8DQ^2}{3r^3}\,.
 \ee
If $DQ^2>0$ and $-M^3<9D/4Q^2$, the metric (\ref{worm}) is regular
for $r>0$, and is actually geodesically complete, as can be checked
by the radial coordinate transformation $r = (8DQ^2/3)\rho^{-2}$,
leading to the behavior
 \be\lb{worm2}
\dd s^2 \simeq \left(\fr{8DQ^2}{3}\right)^2\rho^{-6}(\dd
t-2N\cos\theta \dd \varphi)^2+ 4\dd \rho^2 +
\rho^2(\dd\theta^2+\sin^2\theta \dd \varphi^2)
 \ee
near $\rho\to\infty$ ($r\to0$). Thus the spacetime (\ref{worm}) is a
wormhole interpolating between the two asymptotically flat regions
$r\to0$ and $r\to\infty$ where the curvature invariants
\ba
R&=&-\frac{144 D Q^2 r [2 D Q^2+3 r^2 (M+r)]}{[8
D Q^2+3 r^2 (2 M+r)]^3}\,,\nn\\
R^{\mu\nu\rho\si}R_{\mu\nu\rho\si}
&=&\frac{2592 r^2}{[8 D
Q^2 + 3r^2(2M+r)]^6}\, [608D^4Q^8 - 32D^3Q^6r^2(5M+24r)\\
&& +24D^2Q^4r^4(37M^2+40Mr+30r^2) + 180DMQ^2r^8 + 27M^2r^{10}]\nn\\
\ea
vanish. The dilaton again develops a singularity at a finite
distance.

2) The relations
 \be\lb{r2}
N=[\epsilon(D-M)-\epsilon' \sqrt{2}Q]/2,\quad
A=[\epsilon(D-M)+\epsilon' \sqrt{2}Q]/2, \quad \sqrt{2}P=\epsilon'
(M+D)
 \ee
provide another, less obvious solution to Eqs. (\ref{fc}) and
(\ref{BPS}). The corresponding vectors $\pmb\mu, \pmb\nu$ are given
by
\begin{equation}\label{w4}
    \pmb{\mu} = (-\epsilon' \sqrt{2}Q,\,-\sqrt{2} Q,\, M-D)\,,\quad
    \pmb{\nu}=(M+D,\,\epsilon' (M+D),\,-\epsilon(M-D))\,.
\end{equation}
The target space coordinates read
\begin{align}
& f^{-1}= 1+2 M \tau + \alpha\beta\tau^2(1+\beta\tau/3) \,,\nn\\
& \chi= \epsilon\{1-f[1+\alpha\tau(1+\beta\tau)]\}\,,\nn\\
& \e^{2\phi} = 2[1+(M+D)\tau] - f^{-1}(1+v^2)\,,\nn\\
& \kappa = -\epsilon\e^{-2\phi}[1+(\alpha+2\beta)\tau
-f^{-1}(1-uv)]\,,\label{w5}\\
& v = \epsilon\epsilon'[1-f(1+\alpha\tau)(1+\beta\tau)]\,,\nn\\
& u= \epsilon'[1-f(1+\beta\tau)]\,,\nn
\end{align}
with
 \be
\alpha \equiv M + D - \epsilon\epsilon'\sqrt{2}Q\,, \quad \beta
\equiv M - D\,.
 \ee
As in the case of the representative 1, dualization again leads to a
metric of the form (\ref{worm}), with $f^{-1}(r)$ a cubic function
of $\tau=1/r$ for $D\neq M$ (the solution with $D=M$ belongs to the
strongly degenerate subcase 2), corresponding to a geodesically
complete wormhole spacetime.

\subsection{Non-degenerate case}

In the case $\det B\neq 0$, the matrix $B$ is no longer nilpotent,
so that the matrix exponential in (\ref{ngalf}) does not reduce to a
polynomial in $\tau$. In order to evaluate it, we will make use of the
Lagrange formula
\begin{equation*}
    \e^{B\tau} = \sum_{k=1}^{4}\e^{p_{k}\tau}\prod_{j\neq k}\frac{B-p_j}{p_k-p_j}\,,
\end{equation*}
where $p_j$ are the eigenvalues of $B$, which from (\ref{s2}) are
the four roots of $-\det B=4\pmb{\la}^2$. Contrary to the case of
Lorentzian EMDA, the $SO(2,1)$ norm of the vector $\pmb{\la}$ is
indefinite and so $\det B$ may be positive or negative.

\paragraph*{\pmb{1) $\det B<0$.}} It is convenient to normalize $\tau$ so that $\det B=-1$
(the general case may be recovered by a rescaling of the charges and
an inverse rescaling of $\tau$). The eigenvalues of $B$ are $p_j=\pm
1,\;\pm \ii$, leading to
\begin{equation}\label{nd1}
2\e^{B\tau} = (\cosh\tau + \cos\tau)I + (\sinh\tau + \sin\tau)B +
(\cosh\tau - \cos\tau)B^2 + (\sinh\tau - \sin\tau)B^3\,.
\end{equation}
The corresponding target space coordinates are
\begin{align}\label{nd2}
f^{-1}&= [1/2+G_1]\cosh\tau + [1/2-G_1]\cos\tau +
[M+H_1(d_-,d_+,M)]\sinh\tau +
[M-H_1(d_-,d_+,M)]\sin\tau \,,\nn\\
f^{-1}\chi&=- [N-H_1(d_-,-d_+,N)]\sinh\tau-[N+H_1(d_-,-d_+,N)]\sin\tau \,,\nn\\
f^{-1}v&= G_{2+}(\cosh\tau-\cos\tau) +[Q/\sqrt{2}+H_{2+}]\sinh\tau +
[Q/\sqrt{2}-H_{2+}]\sin\tau\,,\nn\\
f^{-1}u&=G_{2-}(\cosh\tau-\cos\tau)+[P/\sqrt{2}-H_{2-}]\sinh\tau
+[P/\sqrt{2}+H_{2-}]\sin\tau\,,\\
\e^{2\phi}&=[1/2-G_1]\cosh\tau + [1/2+G_1]\cos\tau +
[D+H_1(m_+,m_-,D)]\sinh\tau + [D-H_1(m_+,m_-,D)]\sin\tau -f^{-1}v^2\,,\nn\\
\kappa\e^{2\phi}&=[A-H_1(-m_+,m_-,A)]\sinh\tau +
[A+H_1(-m_+,m_-,A)]\sin\tau -f^{-1}uv\,,\nn
\end{align}
where we have defined
\begin{align*}
& G_1 = m_+m_--d_+d_-\,, \quad G_{2\pm} = \frac1{\sqrt2}[(m_+\pm d_-)q_+ \pm (m_-\pm d_+)q_-]\,,\\
& H_1(x,y,z))=xq_+^2 + yq_-^2 -4xyz\,,\\
& H_{2\pm}=\frac1{\sqrt2}[2m_+d_-q_+ \pm 2m_-d_+q_- \mp(q_+\pm
q_-)q_+q_-]\,.
\end{align*}

\paragraph*{\pmb{2) $\det B>0$.}} Normalizing $\tau$ so that $\det B=+4$, the eigenvalues of $B$ are
$p_j=\pm(1\pm\ii)$, leading to \cite{bps}
\begin{equation}\label{nd3}
2\e^{B\tau} = 2g_1I + 2g_+B + g_2B^2 + g_-B^3\,,
\end{equation}
with
 \be
g_1=\cosh\tau \cos\tau\,,\quad g_2=\sinh\tau \sin\tau\,,\quad
2g_{\pm}=\cosh\tau \sin\tau \pm \sinh\tau \cos\tau\,.
 \ee
The target space coordinates are
\begin{align}\label{nd4}
f^{-1}&=g_1 + 2Mg_+ + G_1g_2 + H_1(d_-,d_+,M)g_- \,,\nn\\
f^{-1}\chi&= - 2Ng_+ +  H_1(d_-,-d_+,N)g_- \,,\nn\\
f^{-1}v&=  \sqrt{2}Qg_+ + G_{2+}g_2 + H_{2+}g_- \,,\nn\\
f^{-1}u&= \sqrt{2}Pg_+ + G_{2-}g_2 - H_{2-}g_- \,,\\
\e^{2\phi}&= g_1 + 2Dg_+ - G_1g_2 + H_1(m_+,m_-,D)g_- - f^{-1}v^2\,,\nn\\
\kappa\e^{2\phi}&= 2Ag_+ - H_1(-m_+,m_-,A)g_-  - f^{-1}uv\,.\nn
\end{align}

In both the above solutions the functions $f^{-1}$ and $\e^{2\phi}$
oscillate and have an infinite number of simple roots for generic
values of the parameters. It is easy to show that the roots $\tau_i$
of the scale factor $f^{-1}$ mark curvature singularities through
which the geodesics cannot be prolongated. Thus the physical
solution must  either lie in the interval $(0<\tau<\tau_1)$ between
the infinity $\tau=0$ and the lowest root, or between two
neighboring roots, $(\tau_i<\tau<\tau_{i+1})$. Only in the first
case the solution is ALF and extremal by construction, so we can
choose it as candidate instanton. However the corresponding on-shell
action, given by (\ref{acdel}) with the upper limit $\tau=\infty$
replaced by $\tau=\tau_1$, is divergent. This solution therefore
cannot be accepted as instanton. It is interesting to note that,
although it saturates the asymptotic no-force bound, it is  not
supersymmetric.

\setcounter{equation}{0}
\section{Discussion of the solutions: exceptional asymptotics}

In this section we shall present the most relevant examples of null
geodesic solutions with the various exceptional asymptotics
behaviors outlined at the end of Sect. 4, without entering into a
detailed discussion of all the possible solutions.

\subsection{Case E1}

In this ALE case, the natural background is flat four-dimensional
Euclidean space \cite{Hunter:1998qe}, with $f_0 = \tau^{-1} =r$.
The regularized trace $[k]$ of the extrinsic curvature of $\pa\cE$
again vanishes, so that the net action $S_{\rm inst} = S_1 + S_2$ is
now given by
 \be\lb{Sale}
S_{\rm inst} = \frac{\beta}4\left(f_0^{-1}\dot{f}_0\Big|_{\tau=0}
-\left[\frac{\dot\Delta}{\Delta}\right]_{\tau=0}^{\tau=\infty}
\right) =  \frac{\beta}2\left(\dot{\phi}(0)
-\frac12\frac{\dot\Delta}{\Delta}\Big|_{\tau=\infty}\right)\,,
 \ee
where we have used the ALE condition $\lim_{\tau\to0}(f-f_0)=0$. For
all degenerate solutions, the contribution of the second term in
(\ref{Sale}) will vanish just as in the ALF case, so that the instanton
action will simply be proportional to the dilaton charge. The matrix $B$ is
now replaced by $B'_1$ given in (\ref{B1}). The dualized one-forms
are given by (\ref{dualcon}) with $-b$ replaced by the lower
left-hand block of $B'_1$, leading to
 \be
\omega = -(B'_1)_{31}\cos\theta\dd\varphi =
\mp2m_{\mp}\cos\theta\dd\varphi
 \ee
for $\tau = 1/r$.

We first discuss the strongly degenerate case. Applying the
transformation (\ref{transf}), with the transformation matrix $K$
given by (\ref{k1}), to the matrix (\ref{ML}) leads to the target
space potentials for the ALE asymptotics:
 \ba\lb{potlin1}
f^{-1} &=& 2m_{\mp}\tau\,, \quad \chi = \mp f\,,  \nn\\
v &=& \frac{q_{\pm}}{2m_{\mp}}\,, \quad u = \pm v\,, \\
\e^{2\phi} &=& 1 + 2D\tau - f^{-1}v^2\,, \quad \kappa =
\pm(1-\e^{-2\phi} - 2d_{\mp}\tau \e^{-2\phi})\,. \nn
 \ea
The constant electromagnetic potentials can be gauged away to  $q_{\pm}=0$,
implying $m_{\pm}= 0$. For the choice $M = 1/4$ (consistent with the ALE
normalisation (\ref{fr})), this leads to the solutions
 \be
f = \mp\chi = \tau^{-1}\,, \quad \e^{2\phi} = 1 + 2D\tau\,, \quad
\kappa = \pm (1-\e^{-2\phi})\,,
 \ee
in the subcase 1, and
 \be
f = \mp\chi = \tau^{-1}\,, \quad \e^{2\phi} = 1 + 2D\tau\,, \quad
\kappa = \mp (1-\e^{-2\phi})\,,
 \ee
in the subcase 2. These are the extremal dilato-axionic instantons
\cite{dilaton-axion} with self-dual scalar fields on a flat
four-dimensional metric, a prototype of D-instantons. The
generalisation to a multicenter harmonic function
 \be
\tau = \sum\limits_{i=1}^s\frac1{|\rr-\rr_i|}
 \ee
leads to non-trivial instanton solutions of the gravitating
dilaton-axion system with a regular metric, namely flat space
(\ref{4eucl}) for $s=1$, the Eguchi-Hanson metric for $s=2$ and lens
spaces for higher $s$ \cite{Eguchi:1980jx}.

The weakly degenerate case leads to dyonic ALE instantons,
generalizing the above solutions. Simple solutions can be obtained
in the case of the representative 1. In the subcase $\epsilon_2 =
-1$, $\epsilon_1 = \mp1$ ($\epsilon = \epsilon' = \mp1$), one
obtains for the choice $M=1/4$
 \ba
f &=& \mp \chi = \tau^{-1}\,, \quad v = \mp u = Q\tau\,, \nn\\
\e^{2\phi} &=& 1 + 2D\tau - \frac{Q^2}3\tau^3\,, \quad \kappa =
\mp(1-\e^{-2\phi})\,.
 \ea
For the boundary action (\ref{acdel}) one obtains, as in the case of
the strongly degenerate dilaton-axion instanton,
 \be
S_{\rm inst} = \frac{\beta D}2 = 2\pi D\,,
 \ee
with $\beta=4\pi$ the period of the angular coordinate $\eta$,
consistent with Rey's Bogomolnyi result \cite{dilaton-axion}.

In the subcase $\epsilon_2 = +1$, $\epsilon_1 = \mp1$ ($\epsilon =
-\epsilon' = \pm1$), one obtains
 \ba
f^{-1} &=& 4M\tau + \frac{16}3DQ^2\tau^3\,, \quad \chi = \mp f\,, \nn\\
v &=& 2Q\tau f(1+2D\tau)\,, \quad u = \pm 2Q\tau f(1-2D\tau)\,,\nn\\
\e^{2\phi} &=& \tau f(1 +
2D\tau)\left(4M-4Q^2\tau-\frac83DQ^2\tau^2\right) \,,\\  \kappa &=&
\mp4\tau^2 f\e^{-2\phi}\left(2MD+Q^2 -
\frac{4}3D^2Q^2\tau^2\right)\,. \nn
 \ea
For $M=1/4$ the metric, of the form (\ref{worm}), is a wormhole
interpolating between an ALE behavior for $r\to\infty$ and the
conical ALF behavior (\ref{worm2}) for $r\to0$ ($\rho\to\infty$).
At spatial infinity, the dilaton behaves as
 \be
\phi \simeq 1 + 2(D-2Q^2)\tau \quad (\tau\to0)\,,
 \ee
and the action is given by (\ref{acE2}) where $D$ is replaced by
that the effective dilaton charge $D-2Q^2$.
In both this and the preceding subcase, the dilaton develops a
singularity at a finite distance. The exceptional, non ALE
possibility $M=0$ leads to a negative definite $\e^{2\phi}$.

\subsection{Case E2}

In the strongly degenerate case, transforming (\ref{ML}) by
(\ref{transf}) with the transformation matrix (\ref{k2}) leads to
the exceptional ALF solution
 \ba\lb{sdE2}
f^{-1} &=& 1 +2M\tau\,, \quad \chi = \frac{N}{M}(f-1)\,,  \nn\\
v &=& q_{\mp}\tau f\,, \quad u = \pm q_{\pm}\tau f\,, \\
\e^{2\phi} &=& \tau f\left[2d_{\mp} +
(4Md_{\mp}-q_{\mp}^2)\tau\right]\,, \quad \kappa =\pm
\e^{-2\phi}f\left[1+2M\tau - q_{\pm}q_{\mp}\tau^2\right]\,. \nn
 \ea

If $d_{\mp} \neq 0$ ($d_{\mp} = 0$ leads to a negative definite
$\e^{2\phi}$), the three-form associated with the axion field is,
asymptotically,
 \be
H \simeq d_{\mp} \,(\dd t\wedge\dd\theta\wedge\sin\theta\dd\varphi)
\quad (r \to \infty)\,,
 \ee
so that the one-form and three-form contributions to the action
(\ref{acH}) are both linearly infra-red divergent. This divergence
is similar to that of the bare (unregularized) purely gravitational
action, suggesting that it can be regularized according to
(\ref{GHax}).  This does not modify the value of the regularized
action for the ALF instantons of Sect. 6 or for the ALE instantons
of case E1, for which the background has a vanishing axion field. In
the present case, we choose as background the solution (\ref{sdE2})
with $M=|N|$ (self-dual Taub-NUT instanton metric) and $Q=P=0$. For
this configuration, $\kappa_0 = \pm \e^{-2\phi_0}$, leading to
$\e^{4\phi_0}\kappa_0\dot\kappa_0 =-2\dot\phi_0$, so that the total
regularized action becomes
 \be
S_{\rm inst} = \frac{\beta}4
\left(\left[\frac{\dot{f}}{f}-2(\dot\phi -
\dot\phi_0)\right]_{\tau=0}^{\tau=\infty}-2|N|\right)\,.
 \ee

For the solution (\ref{sdE2}),
 \be
\dot\phi \simeq \frac1{2\tau} - \frac{q_{\mp}^2}{4d_{\mp}} \quad
(\tau\to0)\,.
 \ee
while $\dot\phi(\infty)=0$, leading to the value of the boundary
action
 \be\lb{acE2}
S_{\rm inst} =
\frac{\beta}2\left[M-|N|-\frac{q_{\mp}^2}{4d_{\mp}}\right]\,.
 \ee
In the subcase 1 with $\epsilon=\mp1$, the solution
 \ba\lb{E21}
f^{-1} &=& 1 +2M\tau\,, \quad \chi = \pm(f-1)\,, \nonumber \\
v &=& 2Q\tau f\,, \quad u = 0\,, \\
\e^{2\phi} &=& 4\tau f\left[D + (2MD-Q^2)\tau\right]\,, \quad \kappa
=\pm \e^{-2\phi}\,,\nonumber
 \ea
is self-dual Taub-NUT supporting a purely electric field and a
self-dual axidilaton. The dilaton field is regular provided $D\ge
Q^2/2M$, however the action (\ref{acE2}) with $d_{\mp}=2D$ is then
negative unless $P=Q=0$. On the other hand, the subcase 2 can lead
to regular instanton solutions with positive action. For instance,
for the two-parameter family (\ref{2magn}), $d_{\mp} = m_{\mp}$ and
$q_{\mp}^2 = 2(M^2-N^2)$, leading to
 \be
\e^{2\phi} = 2m_{\mp}\tau f\left[1 + m_{\mp}\tau\right]\,,
 \ee
which is positive definite if $m_{\mp}>0$, and to the value of the
action
 \be\lb{Se2+}
S_{\rm inst} = \frac{\beta}4\left(M \mp N - 2|N|\right)\,.
 \ee
A sufficient condition for this to be positive, irrespective of the
sign of $N$, is $M > 3|N|$.

In the weakly degenerate case, the representative 1 with $\epsilon_2
= -1$, $\epsilon_1 = \mp1$ leads to a solution which is also
electric Taub-NUT, but with a singular dilaton,
 \ba
f^{-1} &=& 1+2M\tau\,, \quad \chi = \mp(f-1)\,,\nn\\
v &=& 2Q\tau\,, \quad u = 0\,,\nn\\
\e^{2\phi} &=& 4\tau\left(D-Q^2\tau - \frac{2}3MQ^2\tau^2\right)\,,
\quad \kappa = \pm \e^{-2\phi}\,.
 \ea
Because the weakly degenerate and strongly degenerate instanton
solutions have (for $d_{\mp} \neq 0$) the same asymptotic behavior,
the action is again given by (\ref{acE2}) with $M=|N|$ and
$d_{\mp}=2D$, and is again negative. The representative 1 with
$\epsilon_2 = +1$, $\epsilon_1 = \mp1$ gives
 \ba
f^{-1} &=& 1+2M\tau+\frac83DQ^3\tau^3\,, \quad \chi = \mp(f-1)\,,\nn\\
v &=& 4DQ\tau^2f\,, \quad u = \pm2Q\tau f\,,\nn\\
\e^{2\phi} &=& 4D\tau
f\left(1+2M\tau-\frac43DQ^2\tau^3\right)\,,\nn\\
\kappa &=& \pm
f\e^{-2\phi}\left(1+2M\tau-\frac{16}3DQ^2\tau^3\right)\,,
 \ea
leading to a wormhole metric of the form (\ref{worm}), with
vanishing regularized action
 \be
S_{\rm inst} = 0\,.
 \ee
An example of a weakly degenerate representative 2 solution with a
positive action is obtained from (\ref{r2}) with $Q=0$,
$\epsilon=\pm1$, leading to $d_{\mp}= \mp\epsilon' q_{\mp}/\sqrt2 =
m_{\pm}$, so that the action is again given by (\ref{Se2+}), leading
to
 \be
S_{\rm inst}  = \frac{\beta}8\left(3M-D\pm2|M-D|\right)\,.
 \ee

\subsection{Case E3a}

The strongly degenerate case leads to the exceptional ALE solution
 \ba\lb{sde3a}
f^{-1} &=& 2 m_{\mp}\tau\,, \quad \chi = \mp f\,,\nn\\
v &=& \frac{q_{\pm}}{\sqrt2 m_{\mp}}\,, \quad u = 0\,,\nn\\
\e^{2\phi} &=&
2\left(d_{\pm}-\frac{q_{\pm}^2}{2m_{\mp}}\right)\tau \,,\nn\\
\kappa &=& \mp\e^{-2\phi}\,.
 \ea
The constant electric field can be gauged away to $q_{\pm}=0$. The
regularized action is now, for degenerate solutions,
 \be\lb{ace3a}
S = \frac{\beta}2\left[\dot\phi-\dot\phi_0\right]_{\tau=0}\,,
 \ee
with the solution (\ref{sde3a}) itself as the only possible
background, so that the action vanishes identically.

The weakly degenerate representative 1 with $N=\mp M$, $A = \mp D$,
$P = \pm Q$ (the other possibilities lead to vanishing $f^{-1}$ or
$\e^{2\phi}$) leads to
 \ba
f^{-1} &=& 4M\tau+\frac{16}3DQ^2\tau^3\,, \quad \chi = \mp f\,,\nn\\
v &=& 2\sqrt2Q\tau f\,, \quad u = \mp4\sqrt2 DQ\tau^2 f\,,\nn\\
\e^{2\phi} &=& -8Q^2\tau^2 f\,,\nn\\
\kappa &=& \mp
f\e^{-2\phi}\left(4M\tau-\frac{32}3DQ^2\tau^3\right)\,.
 \ea
However, the dilaton field is negative definite. The regularized
action (with $D=0$ as background) again vanishes.

\subsection{Case E3b}

We shall discuss only the two limiting cases $\cos\nu=\pm1$
(exceptional ALE) and $\sin\nu=\pm1$ (magnetic linear dilaton):

1) \underline{$\cos\nu=\pm1$}. In the strongly degenerate case we
obtain
 \ba
f^{-1} &=& 2 m_{\mp}\tau\,, \quad \chi = \mp f\,,\nn\\
v &=& 0\,, \quad u = \pm\frac{q_{\pm}}{\sqrt2 m_{\mp}}\,,\nn\\
\e^{2\phi} &=& 2d_{\mp}\tau \,, \quad \kappa = \pm \e^{-2\phi}\,.
 \ea
The constant magnetic field can be gauged away to $q_{\pm}=0$. The
action (\ref{ace3a}) vanishes as in the case E3a.

The weakly degenerate representative 1 with $N=\mp M$, $A = \mp D$,
$P = \pm Q$ leads, for the choice $M=1/4$, to
 \ba
f^{-1} &=& \tau+\frac{16}3DQ^2\tau^3\,, \quad \chi = \mp f\,,\nn\\
v &=& 4\sqrt2 DQ\tau^2 f\,, \quad u = \pm2\sqrt2Q\tau f\,,\nn\\
\e^{2\phi} &=& 4D\tau^2 f\left(1-\frac{8}3DQ^2\tau^2\right) \,,\nn\\
\kappa &=& \pm f\e^{-2\phi}\left(\tau-\frac{32}3DQ^2\tau^3\right)\,,
 \ea
while the representative 1 with $N=\mp M$, $A = \mp D$, $P = \mp Q$
leads (again for $M=1/4$) to
 \ba
f^{-1} &=& \tau\,, \quad \chi = \mp f\,,\nn\\
v &=& \sqrt2 Q\tau\,, \quad u = 0\,,\nn\\
\e^{2\phi} &=& 4D\tau -\frac{2}3Q^2\tau^3\,, \quad \kappa = \pm
\e^{-2\phi}\,.
 \ea
In both cases, the action again vanishes.

2) \underline{$\sin\nu=\pm1$}. In the strongly degenerate case,
\begin{align}
& f^{-1}=(M+D\mp\sqrt{2}P)\tau\,,\quad \chi=-\frac{N+A}{M+D\mp \sqrt{2} P }
\,,\nn\\
& v=\pm\frac{N-A\pm\sqrt{2}Q}{M+D\mp\sqrt{2} P }\,, \quad
u=\pm\frac{1+(M-D)\tau}{(M+D\mp\sqrt{2} P ) \tau }\,,\nn\\
& \e^{2\phi}=\frac{(M+D\mp\sqrt{2} P)^2-(N-A\pm\sqrt{2}Q)^2}{M+D\mp\sqrt{2} P }\,\tau\,,\\
& \kappa=-\frac{(N-A\pm\sqrt{2}Q)(1+(M-D)\tau)-(M+D\mp\sqrt{2}P)
(N+A)\tau}{[(M+D\mp\sqrt{2} P)^2-(N-A\pm\sqrt{2}Q)^2]\,\tau }\,. \nn
\end{align}
The constant field $\chi$ can be gauged to zero by the choice
$A=-N$. After taking into account the null geodesic and strong
degeneracy conditions (\ref{BPS}) and (\ref{col1}) ((\ref{charA1})
leads to $\e^{2\phi}=0$), there remains the solution
 \ba
&& f = \pm u =  \frac1{4M\tau}\,, \quad \chi = 0\,, \quad v =
\pm\frac{N}M\,, \nn\\ && \e^{-2\phi}=\frac{M}{4(M^2-N^2)\tau}\,,
\quad \kappa= -\frac{N}M\e^{-2\phi}\,,
 \ea
with vanishing regularized action (again, the only natural
background is the solution itself). The weakly degenerate case leads
to complicated expressions which we shall not give here.

\setcounter{equation}{0}
\section{Multiple harmonic functions}
The instantons listed above were incorporating only one independent
harmonic function (including the multicenter solution in which all
centers have equal charges). However, the construction (\ref{AB})
may be generalized \cite{gc,bps} to the case of several truly
independent harmonic functions $\tau_a,\;\Delta \tau_a =0$, by
replacing the exponent in (\ref{AB}) by a linear superposition
\begin{equation} \label{mupt}
 M = A \exp \left(\sum_a B_a \tau_a\right).
\end{equation}
This solves the field equations (\ref{sigeq}) provided that the
commutators $[B_a, B_b]$ commute with the $B_c$ (for the proof see
\cite{bps}):
\begin{equation} \label{dcom}
[\,[B_a, B_b], B_c] = 0 \,.
\end{equation}
The three-dimensional Einstein equations (\ref{ei}) generalize to
\begin{equation}
R_{ij} = \frac{1}{4} \,\sum_a \sum_b \tr (B_a B_b) \,\nabla_i\tau_a
\nabla_j\tau_b \,,
\end{equation}
so that the three-space is Ricci flat if the matrices $B_a$ satisfy
\begin{equation} \label{bal}
\tr (B_a B_b) = 0 \,.
\end{equation}
It follows from the above that the number of independent harmonic
functions on which an extremal solution of the form (\ref{mupt}) may
depend is limited by the number of independent mutually orthogonal
null vectors of the target space. As discussed in \cite{bps}, for a
locally Minkowskian target space with signature ($+p,-q$) the
maximum number of independent null vectors is $\inf (p, q)$. So in
the present case of Euclidean EMDA, BPS solutions depending on three
harmonic functions (as opposed to only two for Lorentzian EMDA
\cite{bps}) are possible in principle.

We show in Appendix D that, in the case of Euclidean EMDA, the
double commutation relations (\ref{dcom}) together with the
Ricci-flatness conditions (\ref{bal}) imply the apparently stronger
commutation relations\footnote{This is not the case e.g. for
Lorentzian Einstein-Maxwell gravity, where the only linearly
independent matrices satisfying (\ref{dcom}) do not commute but
anticommute \cite{bps}.}
\begin{equation} \label{com}
[B_a, B_b] = 0 \,.
\end{equation}
In that case, differentiation of (\ref{mupt}) yields
 \be
M^{-1}\nabla M = \sum_a B_a\nabla\tau_a\,,
 \ee
so that both the expressions (\ref{dualcon}) for the dualized
one-forms and (\ref{acdel}) for the boundary action generalize to
linear superpositions.

Consider two matrices $B(\m,\n)$ and $B'(\m',\n')$ of the form
(\ref{form}) where $\m,\n,\m',\n'$ are any four $SO(2,1)$ vectors.
Then
\begin{equation*}
    \tr (BB')=4(\m\cdot\m'-\n\cdot\n'),\,\quad
    [B,B']=2\begin{pmatrix}
    \mathcal{A}& \mathcal{B}\\
    \mathcal{B}&\mathcal{A}
    \end{pmatrix}\,,
\end{equation*}
where
\begin{align*}
& \mathcal{A}=(\m\wedge\m'-\n\wedge\n')^0\si_2\,,\\
& \mathcal{B}=(\m\cdot\n'-\n\cdot\m')\si_0 -
(\m\wedge\m'-\n\wedge\n')^3\si_1 +
(\m\wedge\m'-\n\wedge\n')^1\si_3\,.
\end{align*}
Thus, the conditions $\tr B^2=0$, $\tr B'^2=0$, $\tr (BB')=0$ and
$[B,B']=0$ lead to the following four scalar equations and one
vector equation
\begin{align}
\label{mh1}& \m^2=\n^2\,,\quad \m'^2=\n'^2\,,\quad \m\cdot\m'=\n\cdot\n'\,,\\
\label{mh2}& \m\cdot\n'=\n\cdot\m'\,,\quad
\m\wedge\m'=\n\wedge\n'\,.
\end{align}

The analysis of this system (see Appendix D) reveals that only
degenerate matrices are allowed, non-degenerate matrices leading
uniquely to one-potential solutions described in the previous
section. There are three two-potential classes of solutions. The
first, with two strongly degenerate subcase 1 generators such that
$\m'\pm\n' = \m\pm\n = 0$ may be directly extended to
three-potential solutions. In the second class, the two generators
belong to the lightlike sector of the strongly degenerate subcase 1,
with $\m'\pm\n' = \m\mp\n = 0$, $\m'\mp\n' \propto \m\pm\n$. In the
third class, one generator $B$ is weakly degenerate, and the other
strongly degenerate generator $B'$ is proportional to $B^3$. We
discuss these three classes in turn.

\subsection{Three-potential class}

The product of any two matrices $B_a$ with $\n_a=\pm \m_a\; \forall
a$ is identically zero, so that (\ref{mupt}) with any number of such
matrices will lead to an extremal solution. However this number is
limited by the number (three) of linearly independent vectors
$\m_a$, leading to three-potential solutions. In the ALF case, these
are
\begin{equation}\label{s6}
M = \eta[I + B_1(\m_1,\pm \m_1)\tau_1 + B_2(\m_2,\pm \m_2)\tau_2 +
B_3(\m_3,\pm \m_3)\tau_3]\,,
\end{equation}
with $\tau_1$, $\tau_2$, $\tau_3$ three independent multimonopole
harmonic potentials. One may choose the three vectors $\m_a$ to
control independently the gravitational, dilato-axionic and
electromagnetic fields,
\begin{equation}\label{s7}
\m_1 = (\pm M,\, 0, \, M)\,, \; \m_2 = (\pm D,\, 0, \, -D)\,, \;\m_3
= (0,\, -\sqrt2Q, \, 0)\,,
\end{equation}
where the first two vectors are null and the third is spacelike,
leading to a solution depending on the three charges $M$, $D$ and
$Q$,
\begin{align}
& f=1\pm \chi=(1+2M\tau_1)^{-1}\,, \nn  \\
& \e^{-2\phi}=1\pm \kappa = [1 + 2D\tau_2 -
2Q^2f(\tau_1)\tau_3^2]^{-1}
\lb{e2phi3}\,,\\
& v = \mp u = \sqrt2Qf(\tau_1)\tau_3\,. \nn
\end{align}
The general solution, which generalizes the one-potential strongly
degenerate subcase 1 solutions, is a linear superposition of
arbitrarily centered self-dual Taub-NUT metrics. The self-dual
scalar and Maxwell sectors are determined by independent charges
$D,\;Q$ and independent multimonopole harmonic functions. Eq.
(\ref{e2phi3}) may also be written
 \be\lb{del3}
\Delta \equiv f^{-1}\e^{2\phi} = (1+2M\tau_1)(1+2D\tau_2) -
2Q^2\tau_3^2\,.
 \ee
Assuming the harmonic potentials to be normalized so that $\tau_i = \tau
+ \rho_i$, where $\tau(r) = 1/r$ and the $\rho_i$ contain only higher
harmonics, we find again the action to be given by (\ref{acsd}).

Eq. (\ref{del3}) shows that the dilaton field will unavoidably
develop singularities near the centers of $\tau_3$, unless these are
also centers of $\tau_1$ and $\tau_2$. Thus for dilaton regularity
one must have $\tau_1=\tau_3+\tau'_1$, $\tau_2=\tau_3+\tau'_2$,
where $\tau'_1$, $\tau'_2$ and $\tau_3$ are three independent
multicenter harmonic functions (\ref{multi}). In terms of these new
potentials, the three-potential extremal solution
\begin{align}
& f^{-1}=1+2M(\tau'_1+\tau_3) \,, \quad f^{-1}v =
\sqrt2Q\tau_3\,,\nn\\
& f^{-1}\e^{2\phi} = (1+2M\tau'_1)(1+2D\tau'_2) +
2[M+D+2MD(\tau'_1+\tau'_2)]\tau_3 + 2(2MD-Q^2)\tau_3^2
\end{align}
(together with the corresponding dual fields) is regular for
$M\ge0$, $D\ge0$ and $Q^2\le2MD$.

From these ALF solutions, three-potential solutions with exceptional
asymptotics (ALE or cases E2, E3a or E3b) may be generated via the
transformations (\ref{transf}) with the appropriate transformation
matrices $K$ given in Appendix B. The general three-potential ALE solution
is, for $M = 1/4$,
 \ba\lb{3ale}
f^{-1} &=& \tau_1\,, \quad \chi = \mp f\,,  \nn\\
v &=& 2Q\frac{\tau_1}{\tau_3}\,, \quad u = \pm v\,, \\
\e^{2\phi} &=& 1 + 2\frac{D\tau_1\tau_2-2Q^2\tau_3^2}{\tau_1}\,, \quad \kappa =
\pm(1-\e^{-2\phi})\,. \nn
 \ea
For a monopole potential $\tau_1 = 1/r$, this corresponds to independent
multicenter self-dual Maxwell and axidilaton fields living on
four-dimensional Euclidean space. The instanton action is again proportional
to the net dilaton charge, while the net electric and magnetic charges vanish.
Similar configurations also exist for a multicenter potential $\tau_1$, with
the Euclidean space replaced by Eguchi-Hanson or lens spaces. The
three-potential exceptional ALF solution (case E2)
 \ba
f^{-1} &=& 1 +2M\tau_1\,, \quad \chi = \pm(f-1)\,, \nonumber \\
v &=& 2Q\tau_3 f\,, \quad u = 0\,, \\
\e^{2\phi} &=& 4f\left[D\tau_2 + (2MD\tau_1\tau_2-Q^2\tau_3^2)\right]\,,
\quad \kappa = \pm \e^{-2\phi}\,, \nonumber
 \ea
is the natural generalization of (\ref{E21}), with again a generically
negative action. In the cases
E3a or E3b, where one-potential instantons yield a vanishing regularized
action, three-potential instantons also lead to a vanishing total action.
The proof goes as follows. The function $\Delta = f^{-1}\e^{2\phi}$ is
generically quadratic in the harmonic potentials $\tau_i$ (for instance,
$\Delta = 8(2MD\tau_1\tau_2-Q^2\tau_3^2)$ in the case E3a). After linearizing
the potentials around the background potential $\tau = 1/r$ according to
$\tau_i = \tau + \rho_i$, we find
 \be
\frac{\dot\Delta}{\Delta} = 2r + O(r^2\rho_i)\,.
 \ee
The monopole component $2r$ is cancelled by the background
subtraction, so that the integrand of (\ref{Sb}) will be given by
the dipole component of the $\rho_i(\rr)$ (the higher multipole
contributions vanish for $r\to\infty$). This is odd in $\rr$, and so
leads to a vanishing boundary action after integration on the outer
boundary.

\subsection{Strongly degenerate two-potential class (dipole instantons)}
Choosing for definiteness the strongly degenerate subcase 1 matrix
$B$ such that $\n = \m$ (this can be changed to $\n=-\m$ by
exchanging the two matrices $B$ and $B'$) with $\m$ lightlike,
$\m^2=0$, the matrix elements of $B'$ are related to those of $B$
through $-\n' = \m' = c^{-1}\m$, with $c$ an arbitrary constant. The
solutions of this class thus depend on three parameters (two for the
null vector $\m$, and $c$). Assuming the two harmonic potentials
$\tau$ and $\tau'$ to be normalized to $\tau\simeq\tau'\simeq 1/r$
at infinity, we choose these parameters to be the net charges $M$,
$N$, $Q$ and $P$ constrained by
 \be\lb{mqnp}
2M^2+Q^2=2N^2+P^2\,,
 \ee
and take
 \be
\n=\m=\bigg(\dfrac{PM+QN}{Q+P},\,-\dfrac{Q-P}{\sqrt{2}},\,
\dfrac{QM+PN}{Q+P}\bigg)\,, \quad
-\n'=\m'=\bigg(\dfrac{PM+QN}{Q-P},\,-\dfrac{Q+P}{\sqrt{2}},\,
\dfrac{QM+PN}{Q-P}\bigg)\,.
 \ee
Because $B^2 = BB' = B'^2 = 0$, the exponential (\ref{mupt})
linearizes, and the equations (\ref{potlin}) giving the target space
potentials in the ALF case generalize to
\begin{align}
& f^{-1}=1+m_+\tau + m_-\tau'\,, \quad
f^{-1}\chi = -m_+\tau + m_-\tau'\,,\nn \\
\label{s4} & f^{-1}v=\frac1{\sqrt{2}}\,[q_-\tau+q_+\tau']\,,\quad
f^{-1}u=\frac1{\sqrt{2}}[-q_-\tau+q_+\tau']\,\,, \\
&f^{-1}\e^{2\phi}= 1 +(m_++d_-)\tau + (m_-+d_+)\tau' +
(m_++d_-)(m_-+d_+)\tau\tau' \nn\\
& f^{-1}\kappa\e^{2\phi}= -d_-\tau + d_+\tau' + (m_+d_+  -
m_-d_-)\tau\tau'\,,\nn
\end{align}
where the dilato-axionic charges $d_{\pm}$ are related to the
gravitational and electromagnetic charges by (\ref{fc}). The
relations (\ref{mqnp}) together with (\ref{fc}) imply that the net
charges again satisfy the balance condition (\ref{BPS}).

We choose for the harmonic potentials $\tau$ and $\tau'$ two
monopole potentials $1/|\rr\pm\bm{a}|$. It is convenient to choose
$\bm{a}$ directed along the $z$ axis and to introduce prolate
spheroidal coordinates $(r,\theta,\varphi)$ related to the the
cartesian coordinates $(x,y,z)$ by
 \be
x=\sqrt{r^2-a^2}\sin\theta\cos\varphi,\quad
y=\sqrt{r^2-a^2}\sin\theta\sin\varphi,\quad z=r\cos\theta,\qquad
(r\geq a)\,.
 \ee
In these coordinates, the three-metric is
\begin{align}
h_{ij}\dd x^i\dd x^j & =  \dd x^2 + \dd y^2 + \dd z^2
%\nn\\ &
= \frac{r^2-a^2\cos^2\theta}{r^2-a^2}\,\dd r^2 +
(r^2-a^2\cos^2\theta)\dd \theta^2 + (r^2-a^2)\sin^2\theta \dd
\varphi^2\,,
\end{align}
and the harmonic potentials are
 \be\lb{potmodi}
\tau=\frac1{r+a\cos\theta}\,,\quad \tau'=\frac1{r-a\cos\theta}\,.
 \ee
The four-dimensional metric
\begin{equation}
\dd s^2 = \fr{r^2-a^2\cos^2\theta}{\Sigma}\,(\dd t-\om_{\varphi}\dd
\varphi)^2 + \Sigma\bigg[\frac{\dd r^2}{r^2-a^2} + \dd \theta^2 +
\frac{(r^2-a^2)\sin^2\theta}{r^2-a^2\cos^2\theta}\dd\varphi^2\bigg]\,,
\end{equation}
with
 \be
\om_{\varphi} = \fr2{r^2-a^2\cos^2\theta}\left[N(r^2-a^2)\cos\theta
+ aMr\sin^2\theta\right]\,, \quad \Sigma =
r^2+2Mr-2Na\cos\theta-a^2\cos^2\theta\,,
 \ee
is supported by the axidilaton and electromagnetic potentials
 \ba
&& \e^{2\phi} = 1 + 2\Sigma^{-1}\left[D(r+M)+A(N+a\cos\theta) +
M^2-N^2\right]\,, \nn\\ && \kappa
= 2\e^{-2\phi}\Sigma^{-1}\left[A(r+M)+D(N+a\cos\theta)\right]\,, \\
&& v = \sqrt2\Sigma^{-1}(Qr + aP\cos\theta)\,, \quad u =
\sqrt2\Sigma^{-1}(Pr + aQ\cos\theta)\,. \nn
 \ea
This solution --- the Euclidean counterpart to the Lorentzian
rotating extremal Taub-NUT dyon of \cite{bps} --- has again finite
action (\ref{acsd}). The metric is regular for $|N|\le M$ ($\Sigma$
cannot vanish in this case, owing to $r\ge a$). The dilaton field is
then also regular ($D^2-A^2=M^2-N^2$ leads to $|A|\le D$, so that
$\Sigma\e^{-2\phi}\ge0$ for $r\ge a$). In the limit $M=\pm N$,
$P=\mp Q$, $A=\mp D$ one recovers the one-potential solutions of the
strongly degenerate subcase 1b. A more interesting limiting case is
$N=A=Q=0$, leading to a magnetic solution with monopole charges
$M=D=\pm P/\sqrt2$ and dipole moments $aM$ (gravitational), $aD$
(axionic) and $aP$ (electric). Finally, one can linearly superpose
such solutions, replacing the harmonic potentials (\ref{potmodi}) by
 \be
\tau = \sum\limits_{i=1}^s\frac1{|\rr-\rr_i+\bm{a}_i|}\,, \quad
\tau' = \sum\limits_{i=1}^s\frac1{|\rr-\rr_i-\bm{a}_i|} \,,
 \ee
with arbitrary orientations and magnitudes of the dipoles
$\bm{a}_i$.

Contrary to the three-potential case, dipole instantons with
exceptional asymptotics are not possible. The reason is that the
transformation matrices (\ref{k1}), (\ref{k2}), (\ref{k3a}) involve
a $\pm$ sign. Before applying the transformation (\ref{transf}) to
the dipole ALF solution, one must choose a definite sign and thus a
definite polarity, so that e.g. it is not possible to superpose
self-dual and anti-self-dual ALE instantons.

\subsection{Third class}
The two-potential solutions of this class are of the form
 \be
M = \eta\left[I + B\tau + \frac12B^2\tau^2 + \frac16B^3(c\tau'+\tau^3)
\right]\,,
 \ee
with $B$ weakly degenerate. These can be treated similarly to the weakly
degenerate solutions considered in Sect. 6B in the ALF case, and in
Sect. 7 for the various exceptional asymptotics, so we do not
repeat the analysis here.

\setcounter{equation}{0}
\section{Six-dimensional interpretation}

As shown in \cite{bremda}, EMDA can be regarded as a consistent
truncation of six-dimensional vacuum general relativity (E6). If the
six-dimensional metric with two commuting Killing vectors is
parametrised by
 \be\lb{624}
\dd s_6^2 = \dd s_4^2 + \lambda_{ab}(\dd x^a + \sqrt2A_{\mu}^a\dd x^{\mu})(\dd x^b
+ \sqrt2A_{\mu}^b\dd x^{\mu})\,,
 \ee
with eleven Kaluza-Klein matter fields $\lambda_{ab}$ and
$A_{\mu}^a$ ($a,b=5,6$), the five covariant constraints
\cite{Clement:2002mb}
 \be
\det(\lambda) = 1\,, \quad F_{\mu\nu}^a = -\epsilon^{ab}\lambda_{bc}
\tilde{F}_{\mu\nu}^c
 \ee
reduce the compactified E6 to EMDA, with only six matter fields.

This may be generalized to the case where both $g_{\mu\nu}$ and
$\lambda_{ab}$ have arbitrary signatures. Assume only
 \be\lb{sd}
 F_{\mu\nu}^a = \eta\epsilon^{ab}\lambda_{bc}
\tilde{F}_{\mu\nu}^c\,,
 \ee
where $\eta = \pm1$ (actually, the sign of $\eta$ is irrelevant). It
follows from (\ref{sd}) that \be\lb{sd1} \tilde{F}_{\mu\nu}^a =
\eta\epsilon^{ab}\lambda_{bc} \tilde{\tilde{F}}_{\mu\nu}^c\,. \ee If
$g_{\mu\nu}$ is Lorentzian, $\tilde{\tilde{F}}_{\mu\nu}^c =
-F_{\mu\nu}^c$, so that
 \be
F_{\mu\nu}^a =
-\epsilon^{ab}\lambda_{bc}\epsilon^{cd}\lambda_{de}F_{\mu\nu}^a =
\det(\lambda)F_{\mu\nu}^a\,,
 \ee implying det$(\lambda) = +1$ (which
is the integrability condition for (\ref{sd}). Conversely, if
$g_{\mu\nu}$ is Euclidean, $\tilde{\tilde{F}}_{\mu\nu}^c =
F_{\mu\nu}^c$, and det$(\lambda) = -1$. It follows that the
six-dimensional signature must in all cases be negative. Two cases
will lead to Euclidean $(+++)$ signature after further reduction to
three dimensions:

1) Six-dimensional signature $(---+++)$. The target space for E6
reduced to three dimensions is $SL(4,R)/SO(4)$ =
$SL(4,R)/(SO(3)\times SO(3))$. After reduction relative to two
timelike Killing vectors ($--$) and truncation, this leads to
phantom EMDA with Lorentzian spacetime and target space (after
reduction to three dimensions) $Sp(4,R)/(SO(3)\times SO(2))$. The
embedding of $sp(4,R)$ in $sl(4,R)$ is treated in Appendix A of
\cite{bps}. Using the conventions of that paper, the $SO(3)\times
SO(3)$ is generated by \be (K^0,\Gamma_0^1,\Gamma_0^2) + (\Sigma_0,
\Gamma_1^0,\Gamma_2^0)\,, \ee while only the first four generators
($U_a=1/2(\Sigma_0, \Gamma_1^0,K^0), U_2=1/2\Gamma_2^0)$)  remain
after truncation to phantom EMDA.

The other case has six-dimensional signature $(-+++++)$, with target
space $SL(4,R)/SO(2,2)$ = $SL(4,R)/(SO(2,1)\times SO(2,1))$. Two
reductions to four dimensions (leaving at least three spacelike
directions) are possible:

2) Reduction relative to two spacelike Killing vectors ($++$),
leading to normal EMDA with Lorentzian spacetime and target space
$Sp(4,R)/(SO(2,1)\times SO(2))$. Choosing for generators of the
$SO(2,1)\times SO(2,1)$ the set \be (\Sigma_0,\Sigma_1,\Sigma_2) +
(K^0,K^1,K^2)\,, \ee the first four of these correspond to the
generators ($V_1$,$W_1$,$U_0$,$U_3$) of $SO(2,1)\times SO(2)$.

3) Reduction relative to one timelike and one spacelike Killing
vectors ($-+$). This leads after truncation to Euclidean EMDA with
target space $Sp(4,R)/(SO(2,1)\times SO(1,1))$. Choosing for
generators of the $SO(2,1)\times SO(2,1)$ the set \be
(\Gamma_0^2,\Sigma_1,\Gamma_2^2) + (\Gamma_1^1,\Gamma_1^0,K^2)\,,
\ee the first four of these correspond to the generators
(\ref{eiso}) of $SO(2,1)\times SO(1,1)$.

Thus, reduction of six-dimensional Lorentzian gravity to four
dimensions Euclidean EMDA is implemented by the constraints
 \be\lb{cons}
 F_{\mu\nu}^a = \pm\epsilon^{ab}\lambda_{bc} \tilde{F}_{\mu\nu}^c\,, \quad
\det(\lambda) = -1\,,
 \ee
the last one being the integrability condition for the first six. In
the following we will note $x^{\mu} = (\xi,r,\theta,\varphi)$ the
four-dimensional Euclidean coordinates (with $\xi$ the Euclidean
time), and $x^a = (t,\eta)$ those of the two extra dimensions. The
standard Euclidean EMDA parametrisation corresponds to the upper
sign in (\ref{cons}) and
  \ba\lambda &=& \left(\begin{array}{cc}
-\e^{-2\phi} + \kappa^2\e^{2\phi} & \kappa \e^{2\phi} \\ \kappa
\e^{2\phi} & \e^{2\phi}
\end{array}\right)\,, \nn\\ A^a &=& (A,\;B)\,, \quad F_{\mu\nu}(B) \equiv
\e^{-2\phi}\tilde{F}_{\mu\nu}(A) - \kappa F_{\mu\nu}(A)\,.
 \ea

For instance, in the neutral case $P=Q=0$, the two-potential
solution (\ref{e2phi3}) uplifted to six dimensions leads (after
putting $\dd\eta'\equiv \pm\dd\eta+\dd t$) to the solution
 \ba\lb{6neutr}
\dd s_6^2 &=& -2\dd\eta'\dd t + \e^{2\phi}\dd\eta'^2
+ f(\dd\xi - \omega_i\dd x^i)^2 + f^{-1}\dd{\bf x}^2\,, \nn\\
\e^{2\phi} &=& 1+2D\tau_2\,, \quad f^{-1} = 1+2M\tau_1\,, \quad
\nabla\wedge\pmb{\om} = \pm\nabla f^{-1}\,,
 \ea
in terms of two independent multicenter harmonic functions $\tau_1$
and $\tau_2$. This corresponds to a six-dimensional plane wave
propagating on a four-dimensional multi-Euclidean Taub-NUT bulk. In
the special case $\tau_1=0$, this solution reduces to
 \be
\dd s_6^2 = -2\dd\eta'\dd t + \e^{2\phi}\dd\eta'^2 + \dd\xi^2 + \dd{\bf
x}^2\,,
 \ee
which is the direct product of the $\xi$ axis by the multicenter
``antigravitating'' solution of five-dimensional vacuum general
relativity found by Gibbons \cite{Gibbons:1982ih} (see also
\cite{gc}). Similarly, the ALE two-potential solution (\ref{3ale})
leads to a six-dimensional vacuum metric similar to (\ref{6neutr})
with $f^{-1} = 4M\tau_1$, corresponding to a flat, Eguchi-Hanson or
lens-space four-dimensional bulk. In the monopole case
$f=r=\rho^2/4$, we recover the multicenter metric with flat
Euclidean four-dimensional bulk
 \be
\dd s_6^2 = -2\dd\eta'\dd t + \e^{2\phi}\dd\eta'^2 + \dd{\bf x_{4}^2}\,,
 \ee
where $\e^{2\phi}$, from the footnote \ref{foo}, is harmonic in the
four-space. This is a special case of the uplift to six dimensions
by (\ref{624}) of four-dimensional dilato-axionic multi-instantons
\cite{Dinstanton} with $\e^{2\phi}$ a generic harmonic function in
four dimensions, previously given in \cite{Clement:2002mb}.

The reduction (\ref{624}) and the constraints (\ref{cons}) are
invariant under $SL(2,R)$ transformations, so that in the
six-dimensional context the fields $\e^{2\phi}$ and $\kappa$ are
defined only up to such transformations. This has two consequences:

1) Exceptional asymptotics of $\e^{2\phi}$ can be transformed to
generic ($\phi(\infty)=0$) asymptotics. If $\e^{2\phi} \sim O(\tau)$
for $\tau\to0$, with e.g. $\kappa\sim -\e^{-2\phi}$, then
$$\lambda(\infty) = \left(\begin{array}{cc} 0 & -1 \\ -1 & 0
\end{array}\right)\,.$$
This can be transformed to generic asymptotics $\lambda(\infty) =$
diag$(-1,1)$ by the linear transformation
 \be\lb{transf6}
\hat\lambda = A^T\lambda A\,, \quad A =
\frac1{\sqrt2}\left(\begin{array}{cc} 1 & 1 \\ 1 & -1
\end{array}\right)\,,
 \ee
leading to
 \ba
\e^{2\hat\phi} &=& \frac12[(\kappa-1)^2\e^{2\phi} - \e^{-2\phi}]
\simeq
1 + \frac12\e^{2\phi}\,, \nn\\
\hat\kappa &=& \frac12\e^{-2\hat\phi}[(\kappa^2-1)\e^{2\phi} -
\e^{-2\phi}] \simeq \e^{-2\hat\phi} - 1\,.
 \ea
So exceptional asymptotics of $\e^{2\phi}$ do not lead to new six-
dimensional solutions.

2) Zeroes of $\e^{2\phi}$ are not really relevant. The matrix
$\lambda$ remains regular at a zero of $\e^{2\phi}$ provided
 \be\lb{reg}
\kappa \e^{2\phi} \simeq \pm1 + O(\e^{2\phi})
 \ee
near such a zero; then, a transformation such as (\ref{transf6})
restores a positive $\e^{2\phi}$. This is the case for the strongly
degenerate solutions of type 1, and presumably also for the strongly
degenerate solutions of type 2. This is also the case for the weakly
degenerate representative 1 with $\epsilon_2 = -1$. In the case of
the ALF weakly degenerate representative 1 with $\epsilon_2 = +1$,
 \be
\e^{2\phi} = f(1+2D\tau)(1+2M\tau-2Q^2\tau^2-\frac43DQ^2\tau^3)\,.
 \ee
For $\tau = -1/2D$, $v$ vanishes, leading to $\kappa \e^{2\phi} = -
\epsilon_1$, and the regularity condition is fulfilled. However,
this does not seem to be the case for the other zeroes of
$\e^{2\phi}$, which would correspond to true singularities.

\section{Conclusion}
To summarize, we have presented a detailed investigation of extremal
Euclidean solutions of EMDA theory (one-vector truncation of $D=4,
N=4$ supergravity) using the purely bosonic technique of
constructing extremal solutions as null geodesic curves of the
three-dimensional sigma-model target space. This target space is the
coset $Sp(4,\R)/GL(2,\R)$ which is yet another homogeneous space of
the $Sp(4,\R)$ U-duality group, apart from the previously discussed
$Sp(4,\R)/U(1,1)$ and $Sp(4,\R)/U(2)$ corresponding to time-like and
space-like reductions of Lorentzian EMDA. The new coset
$Sp(4,\R)/GL(2,\R)$ is a six-dimensional homogeneous space with the
signature $+++---$, and thus possesses three independent null
directions. The Euclidean extremal solutions constitute various
isotropic geodesic surfaces of this space which can be further
classified according to the rank of the corresponding matrix
generators. This purely bosonic classification is a priori not
related with the classification of Killing spinors.

Though the derivation of the three-dimensional EMDA sigma model in
the Lorentzian sector has been known for a long time, we had to
reconsider it in the Euclidean case, taking into account previously
ignored boundary terms arising in the dualizations involved. The
bulk sigma-model action vanishes on-shell, so  the instanton action
is given entirely by the boundary terms. To get rid of infra-red
divergences, which are generically present for non-compact spaces,
we used the matched background subtraction method.  For the
three-dimensional boundary action we then obtained a very simple
expression using the matrix formulation of the sigma model.

Dimensional reduction along a compactified time direction
generically leads to solutions with ALF asymptotic structure.
Instantons with exceptional asymptotics were shown to arise in the
case of asymptotically vanishing (inverse) scale factor of this
reduction (ALE instantons) or, in view of the intrinsic duality
between the metric and axidilaton sectors \cite{Gal'tsov:1996gv}, in
the case of asymptotically vanishing exponentiated dilaton, or in
the combination of these two cases. The coset matrices corresponding
to different asymptotic behaviors were shown to be related by coset
isometries.

Our classification scheme for extremal instantons refers to the
algebraic nature of the corresponding matrix generators and involves
the following types: i) nilpotent rank 2 (strongly degenerate), ii)
nilpotent rank 3 (weakly degenerate) and iii) non-degenerate. Inside
each type, further classification is provided according to the
nature of the charge vectors. The solutions, most of which are new,
include single-center and multi-center harmonic functions. The
instanton action is finite for the classes i) and ii), independently
of the possible presence of singularities of the dilaton function at
finite distance from the centers. The case iii) splits into two
subcases depending on the sign of the determinant of the matrix
generator, in both these subcases the scale factor and the dilaton
function develop singularities at finite distance, and the instanton
action is divergent. Although we did not investigate here Killing
spinor equations, we believe that at least the strongly degenerate
solutions are truly supersymmetric, while the non-degenerate are not
(the weakly degenerate case requires further study).

The above classification relates to solutions generated by a single
harmonic function (including the multi-center case). Our method also
allows for the possibility of multiple independent harmonic functions.
Considering the algebra of generators developing null geodesics of
the target space, one generically finds the
compatibility condition demanding vanishing of the triple
commutators of generators. We have shown that in the case of
Euclidean EMDA these imply a stronger condition of vanishing of all
their pairwise commutators, which effectively linearizes the total
generating current. Given the fact that the Euclidean sigma-model
has three independent null directions (contrary to two in the
Lorentzian case) we have shown that there exist solutions generated
up to three independent harmonic functions. A first class of
solutions is three-potential (all of which are possibly
multicenter) and corresponds to a linear superposition of
arbitrarily centered self-dual Taub-NUTs dressed with self-dual
axidilaton and Maxwell fields. The second class is two-potential
(dipole) and includes the Euclidean counterparts of the EMDA rotating
extremal Taub-NUT and IWP solutions, while the third class, also
two-potential, is built from  a nilpotent matrix generator of rank
3 (weakly degenerate).

Apart from some simple extremal solutions, which were previously
known explicitly in the purely scalar ALE sector
\cite{dilaton-axion}, we were able to construct some new scalar ALF
and ALE solutions, such as dilaton-axion dressed Taub-NUT,
Eguchi-Hanson and lens-space instantons. We also found new types of
solutions which are wormholes interpolating between ALF or ALE and
conical ALF spaces. All electrically and magnetically charged
solutions are entirely new except for those which were (or could be)
found by euclideanization of known Lorentzian black hole and/or
IWP-type solutions, which we rederived in our general treatment as
well. The new charged ALE solutions found here include, among
others, purely electric solutions, as well as purely magnetic
instantons with linear dilaton asymptotics.

The last group of results consists in the six-dimensional uplifting
of the four-dimensional EMDA instantons. Since the three-dimensional
U-duality group $Sp(4,\mathbb{R})$ of EMDA is a subgroup of
$SL(4,\mathbb{R})$, which is the U-duality group of vacuum
six-dimensional gravity reduced to three dimensions, it is clear
that any solution of four-dimensional EMDA of the type discussed in
this paper can be interpreted as a solution of six-dimensional
Einstein gravity without matter fields. We present details of this
relationship and give the explicit six-dimensional form for some such
instanton solutions. This uplift demonstrates, in particular, that
zeroes of the dilaton exponent are not really relevant and can be
resolved in the six-dimensional interpretation.

We have described extremal intantons using a purely bosonic method.
Further work is needed to study the Killing spinor equations for
Euclidean EMDA. We also believe that the new types of BPS instantons
found here might give rise to new families of more general non-BPS
EMDA instantons and wormholes, other methods are needed to study
them. Another perspective of investigation of one-dimensional
subspaces of the target space consists in further reduction of the
sigma model to two-dimensions and application of Lax-pair
integration techniques \cite{Chemissany:2011sr}.

\begin{acknowledgments}
D.G. is grateful to the LAPTH for hospitality in May 2010 and July
2011 while parts of this investigation were performed. He also
wishes to thank Murat Gunaydin for invitation to participate in
``Inaugural Workshop on Black Holes in Supergravity and
M/Superstring Theory'' at Penn State (September 2010) and Miriam
Cvetic, Guillaume Bossard and Cl\'ement Ruef for useful comments.
His work was supported in part by the RFBR grant 11-02-01371-a.
 \end{acknowledgments}

\section*{ Appendix A: Phantom EMDA}
\renewcommand{\theequation}{A.\arabic{equation}}
\setcounter{equation}{0}

Similarly to the case of phantom Lorentzian EMDA, treated in
\cite{phantom2}, the action for phantom Euclidean EMDA, corresponding to
a repulsive coupling of the electromagnetic field to gravity, is obtained
from that of normal Euclidean EMDA by the analytical continuation
$\phi \to \phi+\ii\pi/2$. Kaluza-Klein reduction thus leads to the target
space metric
 \be
\dd l^2 = \frac12\,f^{-2}\dd f^2 -
\frac12\,f^{-2}(\dd \chi + v\dd u - u\dd v)^2 - f^{-1}\e^{-2\phi}\dd
v^2 + f^{-1}\e^{2\phi}(\dd u - \kappa \dd v)^2 + 2\dd \phi^2 -
\frac12\,\e^{4\phi}\dd \kappa^2\,,
 \ee
which has the same signature $(+3,-3)$ as the target space metric (\ref{tar4})
of normal Euclidean EMDA, and thus corresponds to the same coset
$Sp(4,\mathbb{R})/GL(2,\mathbb{R})$. However, analytical continuation of the
matrix representative leads to a matrix representative $\ovl{M}$ of the same
form (\ref{ME}) as for normal EMDA, but with the $2\times2$ block matrices
 \be
\ovl{P} = \e^{-2\phi}\left(\begin{array}{cc} f\e^{2\phi}-v^2&
-v\\-v&-1
\end{array}\right)\,, \quad
\ovl{Q}=\left(\begin{array}{cc} vw -\chi & w\\ w &-\kappa
\end{array}\right)\,,
 \ee
which coincide with those of normal Lorentzian EMDA. The antisymplectic matrix
$\ovl{M}$, with the asymptotic behavior
 \be
\eta =
\left(\begin{array}{cc} \sigma_3 & 0 \\
0  & -\sigma_3
\end{array}\right)\,,
 \ee
is related to that of of normal Euclidean EMDA by the $Sp(4,\mathbb{R})$
transformation (\ref{transf}), with
 \be
K =
\frac12\left(\begin{array}{cc} \sigma_0+\sigma_3 & \sigma_0-\sigma_3 \\
-\sigma_0+\sigma_3 & \sigma_0+\sigma_3
\end{array}\right)\,.
 \ee

\section*{ Appendix B: Exceptional asymptotic behaviors}
\renewcommand{\theequation}{B.\arabic{equation}}
\setcounter{equation}{0}
\renewcommand{\thesubsection}{\arabic{subsection}}
\setcounter{subsection}{0}

The asymptotic solution of the sigma model field equations
(\ref{sigeq}) for the matrix $M(\rr)$ given by (\ref{ME}) is
 \be\lb{asol}
M(r) \simeq A(I + Br^{-1})\,,
 \ee
with $A$ a constant symmetric antisymplectic matrix and $B$ a
constant symplectic matrix. The conditions on the $4\times4$ matrix
$A$ imply that it can be written in block form as
 \be
A = \left(\begin{array}{cc} \alpha & \beta \\ \beta^T & \gamma
\end{array}\right)\,,
 \ee
with the $2\times2$ matrices $\alpha$, $\beta$ and $\gamma$
constrained by
 \ba
&& \alpha^T = \alpha \,, \quad \gamma^T = \gamma\,, \\
&& \alpha\beta^T - \beta\alpha = 0\,, \lb{ab}\\
&& \beta^T\gamma-\gamma\beta = 0\,, \lb{bc}\\
&& \beta^2 - \alpha\gamma = 1\,. \lb{b2ac}
 \ea

As discussed in Sect. 4, the generic matrix $A$ may be
gauge-transformed to $\eta$ given by (\ref{etaE}), corresponding to
the ALF asymptotic behavior, except in the three exceptional cases
1) $f^{-1}(\infty) = 0$, 2) $\e^{2\phi}(\infty) = 0$, and 3)
$f^{-1}(\infty) = \e^{2\phi}(\infty) = 0$. We consider these three
possibilities in turn.

\subsection{$f^{-1}(\infty) = 0$, $\e^{2\phi}(\infty) =1$.}

From (\ref{asol}) $f$ rises linearly as $r$, so the finiteness of
$M_{12}$ implies the asymptotic behavior $v \simeq ar +$ constant,
resulting in $f^{-1}v^2 \sim ar$ which conflicts with the finiteness
of $M_{22}$ unless $a=0$. Thus $v(\infty)$ is constant, and both
$\phi$ and $v$ may be gauge transformed to $\phi(\infty)=0$,
$v(\infty)=0$, leading to
 \be
\alpha = \left(\begin{array}{cc} 0 & 0 \\ 0 & 1
\end{array}\right)\,.
 \ee
The constraint (\ref{ab}) leads to $(f^{-1}u)(\infty)=0$, leading to
$u(\infty)=$ constant, which may again be gauge-transformed to 0.
The matrix $\beta=(P^{-1}Q)(\infty)$ is
 \be
\beta = \left(\begin{array}{cc} -(f^{-1}\chi)(\infty) & 0 \\ 0 &
-\kappa(\infty)
\end{array}\right)\,.
 \ee
The constant $\kappa(\infty)$ may be gauge-transformed to 0, while
the constraint (\ref{b2ac}) leads to
 \be
\chi(r) \simeq \mp f(r) + c
 \ee
asymptotically. Gauging the additive constant $c$ to 0, we obtain
$\beta = \pm(1-\alpha)$, and $\gamma = -\alpha$ from the last
equation (\ref{block}), so that finally
 \be
A = \eta'_1 \equiv \left(\begin{array}{cccc}
0 & 0 & \pm1 & 0 \\
0 & 1 & 0 & 0 \\  \pm1 & 0 & 0 & 0 \\ 0 & 0 & 0 & -1
\end{array}\right) \,.
 \ee
This is related to the matrix $\eta$ of (\ref{etaE}) by
 \be \eta'_1 = K_1^T \eta K_1\,, \ee
with
 \be\lb{k1}
K_1 = \left(\begin{array}{cccc} \is & 0 & \pm\is & 0 \\ 0 & 1 & 0 & 0 \\
\mp\is & 0 & \is & 0 \\ 0 & 0 & 0 & 1 \end{array}\right)
 \ee
(actually there is a one-parameter family of such matrices $K_1$).
The isotropy subagebra leaving invariant $\eta'_1$ is obtained from
(2.44) by
 \be
{\rm lie\,}(H_1') = K_1^{-1}{\rm lie\,}(H)K_1
 \ee
(in the present case, $K_1^{-1} = K_1^T$). So the null geodesics
going through the point $\eta_1'$ are
 \be\lb{transfB}
M_1' = \eta_1' \e^{B_1'\tau} = K_1^T \eta \e^{B\tau} K_1 = K_1^TMK_1
\,,
 \ee
with
 \be\lb{B1}
B_1' = K_1^{-1}BK_1 = \left(\begin{array}{cccc} 0 & - q_{\mp} & \pm
2m_{\pm} & \mp q_{\mp} \\ -q_{\pm} & 2D & \mp q_{\mp} & -2A \\ \pm
2m_{\mp} & \mp q_{\pm} & 0 & q_{\pm} \\ \mp q_{\pm} & 2A & q_{\mp} &
-2D
\end{array}\right)\,,
 \ee
where $m_{\pm} \equiv M \pm N$, $q_{\pm} \equiv Q \pm P$ (we have
kept the original parameters $M$, $N$, etc., which are no longer the
physical charges). The classification into three matrix types
(strongly degenerate, weakly degenerate, non-degenerate) is
invariant under this similarity transformation.

\subsection{$f^{-1}(\infty) = 1$, $\e^{2\phi}(\infty) = 0$}

The treatment of this case closely parallels that of case 1, with
the projector $\alpha$ replaced by $1-\alpha$, and leads to
 \be
A = \eta'_2 \equiv \left(\begin{array}{cccc} 1 & 0 & 0 & 0 \\
0 & 0 & 0 & \mp1 \\  0 & 0 & -1 & 0 \\ 0 & \mp1 & 0 & 0
\end{array}\right)
 \ee
(which is obtained from $\eta'_1$ by a reflection relative to the
antidiagonal and a global sign change). This is related to the
matrix $\eta$ by
 \be \eta'_2 = K_2^T \eta K_2\,, \ee
with
 \be\lb{k2}
K_2 = \left(\begin{array}{cccc} 1 & 0 & 0 & 0 \\ 0 & \is & 0 & \mp\is \\
0 & 0 & 1 & 0 \\ 0 & \pm\is & 0 & \is \end{array}\right)\,,
 \ee
leading to
 \be\lb{B2}
B_2' = K_2^{-1}BK_2 = \left(\begin{array}{cccc} 2M & - q_{\mp} & 2N
& \pm q_{\pm} \\ -q_{\pm} & 0 & \pm q_{\pm} & \mp2d_{\pm} \\ -2N &
\pm q_{\mp} & -2M & q_{\pm} \\ \pm q_{\mp} & \mp2d_{\mp} & q_{\mp} &
0
\end{array}\right)\,.
 \ee

\subsection{$f^{-1}(\infty)=0$, $\e^{2\phi}(\infty)=0$}

In this case, $\alpha = 0$, so that the constraint (\ref{b2ac})
reads
 \be\lb{b21}
\beta^2 = 1\,.
 \ee
Both $f^{-1}$ and $\e^{2\phi}$ go to zero as $r^{-1}$, so that one
can choose a gauge such that asymptotically
 \be
\e^{-2\phi} \simeq f^{-1}\,,
 \ee
and $v(\infty)=0$, leading to
 \be
\beta = \left(\begin{array}{cc} -f^{-1}\chi & f^{-1}u \\ f^{-1}u &
-f^{-1}\kappa
\end{array}\right)(\infty)\,.
 \ee
The constraint (\ref{b21}) then leads to
 \be
\chi^2+u^2 \simeq \kappa^2+u^2 \simeq f^2\,, \quad f^{-2}u(\chi +
\kappa) \simeq 0\,,
 \ee
for $r\to\infty$. These are satisfied if either
 \be\lb{e3a}
\chi \simeq \kappa \simeq \mp f\,, \quad u \simeq 0\,,
 \ee
or
 \be\lb{e3b}
\chi \simeq -\kappa \simeq -f\cos\nu\,, \quad u \simeq f\sin\nu\,,
 \ee
with $\nu$ a real constant. These two possibilities lead, up to a
gauge transformation, to $\gamma = 0$, so that
 \be\lb{eta3}
A = \eta'_{3{\rm a}} \equiv \left(\begin{array}{cccc} 0 & 0 & \pm1 & 0 \\
0 & 0 & 0 & \pm1 \\ \pm1 & 0 & 0 & 0 \\
0 & \pm1 & 0 & 0
\end{array}\right)\,,
 \ee
in the case (\ref{e3a}), or
 \be\lb{eta3b}
A = \eta'_{3{\rm b}} \equiv \left(\begin{array}{cccc} 0 & 0 & \cos\nu & \sin\nu \\
0 & 0 & \sin\nu & -\cos\nu \\ \cos\nu & \sin\nu & 0 & 0 \\
\sin\nu & -\cos\nu & 0 & 0
\end{array}\right)\,,
 \ee
in the case (\ref{e3b}).

In the first case, the matrix transforming $\eta$ into $\eta'_{3{\rm a}}$
is
 \be\lb{k3a}
K_{3{\rm a}} = \is\left(\begin{array}{cccc} 1 & 0 & \pm1 & 0 \\ 0 & 1 & 0 & \pm1 \\
\mp1 & 0 & 1 & 0 \\ 0 & \mp1 & 0 & 1
\end{array}\right)\,,
 \ee
leading to
 \be\lb{B3a}
B_{3{\rm a}}' =  \pm\left(\begin{array}{cccc} 0 & 0 & 2m_{\pm} &
-\sqrt2q_{\mp}
\\ 0 & 0 & -\sqrt2q_{\mp} & 2d_{\mp} \\ 2m_{\mp} & -\sqrt2q_{\pm} &
0 & 0 \\ -\sqrt2q_{\pm} & 2d_{\pm} & 0 & 0
\end{array}\right)\,.
 \ee

In the second case, the transformation matrix is
 \be
K_{3{\rm b}} = \is\left(\begin{array}{cccc} 1 & 0 & \cos\nu & \sin\nu \\
0 & 1 & \sin\nu & -\cos\nu \\
-\cos\nu & -\sin\nu & 1 & 0 \\ -\sin\nu & \cos\nu & 0 & 1
\end{array}\right)\,.
 \ee
Simple expressions can be obtained for the corresponding matrix $B$
in the two limiting cases $\sin\nu = 0$ ($\cos\nu = \pm1$) and
$\cos\nu = 0$ ($\sin\nu = \pm1$).

a) $\cos\nu = \pm1$:
 \be\lb{B3b1}
B_{3{\rm b}}' =  \left(\begin{array}{cccc} 0 & -\sqrt2q_{\mp} & \pm2
m_{\pm} & 0
\\ -\sqrt2q_{\pm} & 0 & 0 & \mp2 d_{\pm} \\ \pm2 m_{\mp} & 0 &
0 & \sqrt2q_{\pm} \\ 0 & \mp2 d_{\mp} & \sqrt2q_{\mp} & 0
\end{array}\right)\,.
 \ee

b) $\sin\nu = \pm1$:
\begin{equation}\lb{B3b2}
    B_{3{\rm b}}'=\begin{pmatrix}
 -D+M & \mp(A+N) & -A+N\mp\sqrt{2}Q & \sqrt{2} P\pm(D+M) \\
 \pm(A+N) & D-M & \sqrt{2}P\pm(D+M) & -A+N\mp\sqrt{2}Q \\
 A-N\mp\sqrt{2}Q & -\sqrt{2}P\pm(D+M) & D-M & \mp(A+N) \\
 -\sqrt{2}P\pm(D+M) & A-N\mp\sqrt{2}Q & \pm(A+N) & -D+M
\end{pmatrix}\,.
\end{equation}

\section*{ Appendix C: Multiple null generators}
\renewcommand{\theequation}{C.\arabic{equation}}
\setcounter{equation}{0}

Let us first show that the double commutation relations
(\ref{dcom}), together with the Ricci-flatness conditions
(\ref{bal}) imply the commutation relations (\ref{com}). Consider
the relations (\ref{dcom}) with $a=1$, $b=2$ and $c=1$ or $2$,
 \be
\left[[B,B'],B\right]=0\,, \quad \left[[B,B'],B'\right]=0\,,
 \ee
with $B=B_1$, $B'=B_2$. These equations may be rewritten as
\begin{align}
\label{t10}&
\pmb{\mu}\wedge(\pmb{\mu}\wedge\pmb{\nu'})+\pmb{\mu'}\wedge(\pmb{\nu}\wedge\pmb{\mu})
+\pmb{\nu}\wedge(\pmb{\mu'}\wedge\pmb{\mu})=0\,,\\
\label{t11}&
\pmb{\nu'}\wedge(\pmb{\nu}\wedge\pmb{\mu})+\pmb{\nu}\wedge(\pmb{\mu'}\wedge\pmb{\nu})
+\pmb{\mu}\wedge(\pmb{\nu}\wedge\pmb{\nu'})=0\,,\\
\label{t8}&
\pmb{\mu}\wedge(\pmb{\mu'}\wedge\pmb{\nu'})+\pmb{\mu'}\wedge(\pmb{\nu}\wedge\pmb{\mu'})
+\pmb{\nu'}\wedge(\pmb{\mu'}\wedge\pmb{\mu})=0\,,\\
\label{t9}&
\pmb{\nu'}\wedge(\pmb{\nu'}\wedge\pmb{\mu})+\pmb{\nu}\wedge(\pmb{\mu'}\wedge\pmb{\nu'})
+\pmb{\mu'}\wedge(\pmb{\nu}\wedge\pmb{\nu'})=0\,.
\end{align}
We also have the relations (\ref{bal}) for $a,b=1,2$,
\begin{align}
\label{t2}& \pmb{\mu}^2 = \pmb{\nu}^2\,,\quad \pmb{\mu'}^2 = \pmb{\nu'}^2\,,\\
\label{t3}& \pmb{\mu\cdot\mu'} = \pmb{\nu\cdot\nu'}\,.
\end{align}
Using~\eqref{t2} and~\eqref{t3},~\eqref{t10} may be rewritten as
\begin{equation}
    (\pmb{\nu}^2)\pmb{\nu'}-(\pmb{\mu\cdot\nu'})\pmb{\mu}+(\pmb{\mu'\cdot\nu})\pmb{\mu}
    -(\pmb{\nu\cdot\nu'})\pmb{\nu}+\pmb{\nu}\wedge(\pmb{\mu'}\wedge\pmb{\mu})=0\,,
\end{equation}
or
 \be
a\m = \b\wedge\n\,,
 \ee
with
 \be\lb{defab}
a \equiv \m'\cdot\n - \m\cdot\n'\,, \quad \b \equiv \m\wedge\m' -
\n\wedge\n'\,.
 \ee
Reasoning similarly with eqs. (\ref{t11})-(\ref{t9}), we arrive at
the system
 \be\lb{main}
a\m = \b\wedge\n\,, \quad a\n = \b\wedge\m\,, \quad a\m' =
\b\wedge\n'\,, \quad a\n' = \b\wedge\m'\,.
 \ee

A consequence of (\ref{main}), obtained by iteration, is
 \be
a^2\m = \b\wedge(\b\wedge\m) = (\b^2)\m\,,
 \ee
and a similar equation with $\m$ replaced by $\n$. Excluding the
trivial solution $\m=\n=0$ ($B=0$), these relations give
 \be
a^2 = \b^2\,.
 \ee
We first assume $a\neq0$, implying $\b\neq0$. Taking the wedge
product of the first eq. (\ref{main}) by $\n$, and of the second eq.
by $\m$, and taking into account $\m\cdot\b=\n\cdot\b=0$ (which also
follows from (\ref{main})), we obtain
 \be
a\m\wedge\n = -(\m^2)\b = (\n^2)\b\,,
 \ee
together with a similar primed equation. Because of (\ref{t2}),
these imply
 \be
\m^2 = \n^2 = 0\,, \quad \n = c\m\,, \quad \n' = c'\m'\,.
 \ee
Inserting these two last equations into (\ref{main}), one obtains
$$a\m = c\b\wedge\n = c^{-1}\b\wedge\n\,,$$
and a similar primed equation, so that $c^2 = c'^{\,2} = 1$. Then from
(\ref{defab}) $a = (c-c')\m\cdot\m' = c(1-cc')\m\cdot\m'$, which
vanishes from (\ref{t3}), contrary to our hypothesis.

It follows that $a=0$. Then necessarily also $\b=0$ (if one assumes
$\b\neq0$, then from (\ref{main}) the four vectors $\m$, $\n$,
$\m'$, $\n'$ are all collinear with the same vector $\b$, leading to
$\b=0$). The four equations $a=0$, $\b=0$,
 \be\lb{ab0}
\m'\cdot\n = \m\cdot\n'\,, \quad \m\wedge\m' = \n\wedge\n'
 \ee
are equivalent to $[B,B']=0$.

Now we discuss the system of equations (\ref{t2}), (\ref{t3}),
(\ref{ab0}) for a two-potential BPS solution. Putting $\m_{\pm} =
\m\pm\n$, $\m'_{\pm} = \m'\pm\n'$, this system may be rewritten as
 \ba
&& \m_+\cdot\m_- = \m_+\cdot\m'_- = \m'_+\cdot\m_- = \m'_+\cdot\m'_-
= 0 \lb{scal}\\ && \m_+\wedge\m'_- + \m_-\wedge\m'_+ = 0\,. \lb{vec}
 \ea
Taking successively the wedge product of the vector equation
(\ref{vec}) with the four vectors $\m_+$, $\m_-$, $\m'_+$, $\m'_-$,
we obtain the secondary system
 \ba\lb{sec}
&& (\m_+^2)\m'_- - (\m_+\cdot\m'_+)\m_- = 0\,, \quad (\m_-^2)\m'_+ -
(\m_-\cdot\m'_-)\m_+ = 0\,, \nn\\ && (\m^{'2}_+)\m_- -
(\m'_+\cdot\m_+)\m'_- = 0\,, \quad (\m^{'2}_-)\m_+ -
(\m'_-\cdot\m_-)\m'_+ = 0\,.
 \ea

First assume that none of the vectors $\m_{\pm}$, $\m'_{\pm}$
vanishes. If also all these vectors are non-null, then from the
system (\ref{sec}) $\m'_+$ is proportional to $\m_+$ and $\m'_-$ is
proportional to $\m_-$. In that case, by replacing the original
harmonic potentials $\tau$ and $\tau'$ by suitable linear
combinations of $\tau$ and $\tau'$, one can translate $\m'_+$ or
$\m'_-$ to zero, contrary to our assumption. If one of the vectors,
e.g. $\m_+$ is null, $\m_+^2=0$, then the first and third equations
(\ref{sec}) give also $\m_+\cdot\m'_+ = \m^{'2}_+ = 0$, so that
$\m'_+$ must be proportional to $\m_+$, and again may be translated
to zero by a redefinition of the harmonic potentials $\tau$ and
$\tau'$. So we conclude that at least one of the four vectors
$\m_{\pm}$, $\m'_{\pm}$ must vanish (up to a redefinition of the
harmonic potentials).

Assume that e.g. $\m'_{+}=0$. Then, from Eq. (\ref{vec})
$\m_{+}=c\m'_{-}$. There are two possibilities:

a) \underline{$c=0$}. Then
 \be\lb{}
\m'_{+} = \m_{+} = 0\,,
 \ee
which solves all the equations (\ref{scal}) and (\ref{vec}). Both
matrices $B$ and $B'$ belong to the strongly degenerate subcase 1.

b) \underline{$c\neq0$}. Then from (\ref{scal}) one obtains
$\m_{+}\cdot\m_{-}=0$ and $\m^2_{+}=0$, so that the matrix $B$ is
degenerate and $B'$ belong to the lightlike sector of the strongly
degenerate subcase 1. This can be further divided into three
subcases. In the first ($\m_{-}=0$), $B$ is also in the lightlike
sector of the strongly degenerate subcase 1, with
 \be\lb{}
\m^2_{+}=0\,, \quad \m_{-} = \m'_{+} = 0\,, \quad \m'_{-}=b\m_{+}
 \ee
($b=c^{-1}$).

In the second subcase, $\m_{-}\propto\m_+$, $B$ is
strongly degenerate subcase 2,
 \be\lb{}
\m^2_{+}=0\,, \quad \m_{-} = a\m_+\,, \quad \m'_{+} = 0\,, \quad
\m'_{-}=b\m_{+}\,.
 \ee
However, this second subcase is equivalent to the first, as can be
shown by taking the linear combinations $\tilde{B} = B - (a/b)B'$,
$\tilde{B}' = B'$, leading to  $\tilde{\mu}_+ = \mu_+$ (so that
$\tilde{\mu}_+^2=0$), and $\tilde{\mu}_-=0$.

In the third subcase, $B$ is weakly degenerate,
 \be\lb{}
\m^2_{+} = 0\,, \quad \m_{+}\cdot\m_{-}=0\,, \quad \m'_{+} = 0\,,
\quad \m'_{-}=b\m_{+}\,.
 \ee
Note that in the case of a weakly degenerate matrix $B$,
${\pmb\nu}\wedge{\pmb\lambda} = - (\m^2)\m_+$, $\m\wedge{\pmb\lambda}
= (\m^2)\m_+$,
with $\m^2\neq0$, so that from (\ref{b3b}) one can identify
\be
B' = -\frac{b}{4\m^2}B^3\,.
\ee

%\newpage

\end{document}